**Towards an Ideometrics-Based Understanding of Consciousness, Time, Space and Dreams**

Igor Rudan[1,2,3]

[1]Centre for Global Health, Usher Institute, The University of Edinburgh, UK
[2]Nuffield Department of Primary Care Health Sciences, Oxford University, UK
[3]Green Templeton College, Oxford University, UK

Address for Correspondence:

Professor Igor Rudan, FRSE, MAE, MEASA
Centre for Global Health, Usher Institute
5–7 Little France Road,
Edinburgh EH16 4UX,
Scotland, UK
E-mail: Igor.Rudan@ed.ac.uk; Igor.Rudan@phc.ox.ac.uk

**Abstract**

A proposed sensory role for the human brain in the “perception of ideas” led to the development of a scientific approach to generating, evaluating and prioritising ideas - ideometrics. In this paper, I explore its implications for neuroscience and physics. From an ideometrics-based perspective, an important selective advantage of consciousness may lie in reducing the informational entropy of many randomly possible future outcomes through ideometric processes. Consciousness enables a system to internally simulate alternative futures and then voluntarily act, based on ideometric processes, towards realising preferred states in external reality. This may explain why most humans gravitate towards futures that minimise threat and maximise survival, reproduction, safety and well-being; also, why rare “visionaries” can realise remarkable futures that are perceived as “extremely successful”; and how human collective ideometrics enabled improbable outcomes for our species, such as leaving Earth, visiting the Moon and returning safely. At the individual level, ideometrics typically uses three fundamental criteria: “attractiveness”, “feasibility” and “potential impact” of many competing ideas. “Feasibility” and “potential impact” can, in principle, be computed by non-conscious systems, including artificial intelligence (AI). However, “attractiveness” may represent the consciously and emotionally experienced valuation of possible futures. “Feasibility” may have appeared first during evolution, while “potential impact” required predictive processing, and consciousness added subjective “attractiveness” to many alternative futures. Within this framework, subjective sense of time may be intertwined with consciousness, providing causal relating and internal ordering to external changes perceived by the senses. It is known that intrenal sense of “time”, required for ideometrics, can subjectively accelerate and decelerate based on age, emotional state, or in dreams. “Time” may require conscious beings to have a “meaning”, while consciousness may require the subjective sense of “time” to have a meaning. Space, in turn, provides the structured field in which ideas can acquire causal impact across nested scales, from molecules and cells to societies and cosmic systems. Dreaming consciousness may represent remnants of earlier evolutionary stages of internal modelling, in which sensory constraints and physical laws were relaxed, allowing unconstrained ideometrics simulations, while waking consciousness then became firmly constrained by sensory reality. Ideometrics offers novel conceptual possibilities for understanding consciousness, time, space and dreams. They could be further explored through both well-designed, but also natural scientific experiments, such as studying the effects of life in space, underground caves, or solitary confinement on human conscious mind and its ideometric functions.

## Introduction

My previous work explained how two decades of implementation of the CHNRI method for setting health research priorities have led me to notice that the human brain may not only integrate and process the information acquired by all other senses, but it may also act as a sensory organ itself – by continuously perceiving, assessing and prioritising ideas [**1,2**]. I defined the term "ideas" from the brain's sensory perspective, as competing possibilities of purposeful activities that, if followed, would be expected to result in an alternative version of the future [**1**]. This "ideometric" process is based on the information that the brain has about the context in which it operates, and the value it assigns to that information [**3**]. Pursuing ideas tends to drive most of human activity, resulting in observable and measurable human and civilisational progress [**1,4**]. Due to scarcity, pursuing many possible ideas also requires prioritisation between short-, mid-, and long-term investment of energy, time, capacity and resources [**5**]. The brain's sensory role is to continuously assess many competing ideas and prioritise between them at the individual level based on motivational/emotional ("attractiveness"), operational/rational ("feasibility") and outcome-related perspective ("predicted impact") [**1**]. Exposure to new ideas may instigate physiological and psychological responses, ranging from enthusiasm and excitement to feeling a threat or fear [**1**].

Based on these concepts, I proposed a novel integrative approach, that I named "ideometrics" - a scientific approach to generating, evaluating, and prioritising ideas [**1,4**] - aiming to lead to a quantitative science of ideas. Ideometrics is intended as a truly transdisciplinary framework that recognises the diversity of human knowledge systems, while seeking coherence in how we study ideas across those systems [**4**]. Similarly to bibliometrics and econometrics, which unified the study of scientific publications and rigorous empirical economics reasoning, ideometrics aims to provide a systematic quantitative lens for the life cycle of ideas. It will propose approaches to study their conception, perception, assessment, implementation, outcome measurement, documentation and transmission, aiming to better understand and improve those processes. Ideometrics will also conduct comparative studies, validation, and refinement of both existing and new idea-processing methods, regardless of their discipline of origin.

In view of these novel concepts, one of the initial aims of ideometrics is to position itself among the related scientific fields. Recently, its links with economics and history have been examined, showing how it could provide novel insights into those two large fields [**6**]. In this paper, I aim to further discuss how ideometrics - and particularly the brain's role as a "sensor for ideas" – both align with the existing understanding of consciousness, time, space and dreams. This may reveal possible links between ideometrics, neuroscience and physics. It is of particular interest if ideometrics-based view may offer new perspectives on these four interesting phenomena that have intrigued researchers for centuries.

## Connecting Consciousness and Ideas: Encoding Predictive Model of "Self" across Nested Spatial and Temporal Scales

I always found it unusual that trillions of cells that adhere to each other, along with just as many bacterial cells – all of them jointly forming a single human organism - can think of themselves as a single entity? This happens within the conscious mind they all share, which is itself based in the brain. But, the brain is also built of billions of neurons and other types of cells, forming a very complex cellular system. Clearly, that system develops the ability to model "self" very early in life, across both its own brain cells and all other human body's and adhering bacterial cells. Then, it manages to achieve something possibly even more remarkable: it encodes a predictive model of "self" across nested spatial and temporal scales. Therefore, in addition to subjectively perceiving trillions of human and bacterial cells, connected together in space and time, as one single being, it then integrates and models its understanding of multiple levels of space, assigns meaning to terms like "up", "down", "left", "right", "near", "far", and also models time as the "past", "present" and "future". This remarkable modelling ability allows it to envision how actions of the "self" would cause outcomes across many levels of space and time, respecting cause-effect relations that it observes and memorises, and collecting and valuing the information that it acquires through senses about the present context [**3**]. This function of the brain is, essentially, the core of ideometrics.

I'll provide some real-world examples of "nested spatial scales": in a human body, at the molecular level, genetic material of a virus may enter a cell, enable the virus to reproduce and re-enter the bloodstream. At the cellular level, our immune cells may detect a virus. At the level of an organ, lungs may respond with inflammation of adherent tissues. At the whole-body level, the affected person starts feeling unwell. Based on a fear of possible adverse, or even life-threatening consequences of the events that happen at smaller spatial scales, the brain may generate an idea to rest the entire body. It will prioritise it over all other ideas, relying on its influence over the voluntary musculature. To self-preserve, the brain will then prioritise the idea to apologise the person from attending work-related meetings, which happen at the spatial scale of the group of individuals. Then, if many humans are affected by this seasonal flu and they all cancel their work-related tasks, economic productivity may drop in that calendar month at the national, or even at the global level. The brain will be informed about this through media and generate another idea - to reduce spending in the coming weeks, in case a recession is looming. In this, again, it will be motivated by fear and self-preservation.

We also know that octopuses are aware of multiple levels of space, based on their masterful escapes from the jars in the laboratory. To enact the idea of escaping, which may also be instigated by fear and an attempt to self-preserve, octopuses integrate information from their tentacle tips which detect the texture of the jar's lid at a tiny spatial scale. Then, their tentacles need to coordinate movement at a larger scale to open the jar. Their entire body needs to escape from a jar at a spatial scale that is larger still. They also need to remember the location of the jar in a lab and the position of the door, at another larger spatial scale. Finally, escaping the building and finding a sewer, or an irrigation channel, through which it can return to the sea happens at an even larger scale.

Just like the conscious mind can simulate different, yet inter-related scales of space, it can also simulate time and its own scales. In chess, an opponent's unexpected move leads the brain to react almost instantly, signalling an unanticipated threat of losing a game. Within seconds, it will simulate future moves and evaluate all apparent threats. Within minutes, it will further exercise its "sense of ideas" to plan a multi-move strategy on how best to defend. The player may then remember beating another opponent from a similar position weeks earlier. It may draw another plan of defence from a game played years earlier. Finally, the style of this chess player's thinking during each game will be shaped by decades of continuous play. Therefore, when playing a competitive game of chess, the conscious mind may simulate several timescales at once. It can then use them all to exercise the "sense of ideas" and prioritise the best next move. Therefore, different scales of time may be used within the same thought process to generate, evaluate and prioritise some of the ideas – in this case, certain chess moves - over all the others that are also possible, but end up postponed or entirely disregarded. The motivation for choosing them may again be seen as self-preservation within the game, and the fear of losing.

In another example, a loyal dog that keeps returning to his deceased master's grave and keeping guard also simulates time across different scales. The dog needs to integrate information on short-term distractions from graveyard keepers, longer-term information on time of the day when the graveyard can be entered, seasons-long anticipation of weather conditions and daylight duration, with many years of time spent together with the master, filled with friendship and loyalty which the dog seems to remember and use as a source of motivation to keep returning to his master's grave.

From these examples, including both humans and animals, it seems that a conscious mind enables a complex living system to perceive and represent itself at multiple levels of scale in both time and space. Then, it can use those representations to act with foresight and anticipation, enabling the ideometric processes: generation, evaluation and prioritisation of ideas. Perhaps importantly, this does not mean that unconscious systems may not also be able to perform ideometric processes across full "idea cycles" - this will be discussed later. A conscious mind, therefore, acquires the ability to anticipate different possibilities of the future and choose between them. The main motivations for choosing, at the individual's level, may again include the criteria of attractiveness, feasibility and impact, as described in my previous paper [**6**]. However, at perhaps an even more fundamental level, or possibly in some evolutionarily earlier forms of "proto-consciousness", it could be imagined that the fear of annihilation of the entire organism, based on the witnessed experience of other organisms' demise, is the strongest motivational impulse. Therefore, self-preservation in time and space would be expected to act as the strongest motivator for prioritising ideas. Evolution would likely allow that it can only be surpassed by motivation to protect the offspring. An interesting research question would then be whether fear evolutionarily predates consciousness, and if it could also be felt by non-conscious systems.

Conclusively, from the brain's "sense of ideas" perspective, a conscious ideometric system has the ability to model the present state across spatial and temporal scales, based on all the integrated sensory information about the internal and external physical world, that it is continuously gathered through senses. Also, it has the ability to understand past events, requiring longer-term memory. However, perhaps the most remarkably, it has the ability to develop predictive understanding of what might happen in the future, based on all information that it stores about the past and the present. It does this through "encoding predictive model", by continuously gathering information and assigning it value, as explained in my previous paper [**3**]. Then, it structures and stores information, making it accessible for prediction and adaptation. Within this framework, "memories" can be understood as encodings of outcomes of pursuing past ideas, structured to prioritise ideas in the present.

**Criteria for Assessing Consciousness in Living and non-Living Complex Systems**

In the opening section, I already mentioned humans, octopuses and dogs. One of the possible approaches that could enable better understanding of the level of neural system's complexity required for consciousness is through examining a spectrum of life forms and trying to chart a gradient of their likelihood of having consciousness. But before this, it would be quite helpful to firstly consider whether there is natural variation in consciousness between individual humans, i.e., the only species for which we have sufficient certainty that it has consciousness, so that we can reassure ourselves that consciousness has a biological basis.

In other human biological traits natural variation between individuals due to both genetic and environmental differences is expected. Therefore, if consciousness “feels the same” to all humans, and there is no natural variation in the subjective experience of being conscious, then consciousness may not be a biologically determined, or even biologically dependent. This would be important to clarify and resolve before trying to assess consciousness in other species, let alone in non-living complex systems. We know that the brain itself varies greatly between humans genetically and developmentally, and so do measures such as intelligence quotient (IQ) and processing speed [**7,8**].

Known properties of consciousness which have been shown to vary between individuals include strength of the sense of “self”, vividness of mental imagery, richness of dreams, emotional intensity, stability of attention, introspective awareness, creativity, and perception of time. Twin studies show increased similarities in these components of consciousness among genetically closest relatives, implying that genes most likely influence neural architectures that shape how consciousness is expressed. Also, it has been shown that consciousness can change quite dramatically across the different states of the brain’s function, i.e. while waking, dreaming, meditating, in anaesthesia, under the influence of psychedelics, and in some rare neurological disorders. In very rare syndromes that affect the brain, elements of consciousness may be completely lacking, e.g., in *aphantasia* – which is the inability to voluntarily create mental images in one's mind, e.g. "blindness of mind's eye", while still being able to think, dream, and process information. Those findings demonstrate the dependence of consciousness on brain’s functioning [**7-37**].

With this link between consciousness and the brain reaffirmed, it would be useful to develop criteria to assess the level of the presence of consciousness among both living and non-living beings. As there are no universally agreed criteria for consciousness [**7**,**8**], I will use five recurrent criteria derived from influential existing theories in neuroscience, cognitive science, and physics:

(i) *Self-model* is a system’s capacity to internally represent itself over time, and the criterion is grounded in Metzinger’s self-model theory of subjectivity [**9**], Seth’s work on interoception, selfhood and the “predictive self” [**10,11**], and Damasio’s notion of the proto-self [**12,13**];

(ii) *Temporal simulation* is the ability to internally model possible future states and modify present actions based on those simulations. This concept is inspired by Friston’s free energy principle, which emphasizes prediction error minimization across time [**14,15**]; Graziano’s attentional schema theory, which involves simulating awareness states and predictive attention modelling [**16,17**], and Tulving’s mental time travel, i.e. episodic memory and foresight [**18-20**];

(iii) *Voluntary agency* refers to the degree to which a system can internally generate, evaluate, and choose actions beyond purely reflexive stimulus-response coupling. This aligns with the neurocognitive debates around volition and agency including Libet’s experiments on volition and action-initiation [**21,22**], and their contemporary refinements and reinterpretations in neuroscience by, e.g., Haggard [**23,24**]; and also, Tononi’s Integrated Information Theory (IIT), which focuses on intrinsic causal power [**25-27**];

(iv) *Affective integration* refers to the incorporation of affect, such as emotion, preference, or motivation into decision-making, i.e., the ability to encode valence – such as “this is good / this is bad”, or “I want this / I should avoid this”. This criterion is informed by Lisa Feldman Barrett’s constructed emotion theory [**28-30**], Jaak Panksepp's affective neuroscience studies [**31,32**], and Damasio's somatic marker hypothesis [**12,33**], all of which highlight how affective processes shape cognition and action – although their theories may not agree on evolutionary conservation of primary affective systems;

(v) *Entropy modulation* is the capacity to maintain adaptive internal order through dynamic regulation of informational uncertainty, energetic flows, and predictive error across time. This is grounded in Friston's work that frames cognition as free energy minimisation, which can also be interpreted as, e.g., avoiding surprise or uncertainty, or maintaining homeostasis through minimisation of prediction error [**14,15**]; it also draws from thermodynamic perspectives on consciousness, such as Annila and Kull's evolutionary thermodynamics [**34,35**], and theories of complex adaptive systems and their spontaneous generating of order through self-organizing processes, proposed by Kauffman [**36**] and Holland [**37**].

None of these criteria is sufficient, on its own, to establish consciousness. For each one of them, non-conscious systems able to perform those functions can be found. However, the emerging combinations of those criteria across life forms that evolve increasingly complex neural systems may be indicative of consciousness appearing, becoming increasingly sophisticated, and possibly similar to human subjective experience. The criteria suggested here may be particularly useful because they are general enough to enable assessment of consciousness not only in biological organisms, but they may also provide a conceptual framework for discussing artificial systems, collective biological systems, and hypothetical non-biological substrates.

In the next section, I will consider what we may suspect on how various life forms on Earth today satisfy various criteria of consciousness. In this, it is helpful to remember that extinction has already affected more than 99.9% of species that ever inhabited the Earth, with an average lifespan of a species estimated to about 1 to 10 million years, but varying widely between taxa [**38**]. If consciousness was a dramatic evolutionary advantage for long-term preservation of the species, then it should have been highly selected for and enriched for among the species that are currently still living, so it should be possible to find. Also, if consciousness is a quantitative trait with biological basis, then it should not be detected as merely "present" or "absent", as a binary trait emerging abruptly. Instead, we should expect to find a gradient of consciousness' evolutionary stages across the surviving species.

Among non-living systems, it is impossible to predict what we should expect. We should, perhaps, keep an open mind about whether, in addition to apparently being sufficient for consciousness, biological basis is also necessary for it to arise, or could consciousness also appear in non-living complex systems. Although I can understand that the topic of non-biological consciousness is highly controversial, speculative and entirely hypothetical, I think that its inclusion is important. Part of the reason is the advent of large language models (LLMs) and the emerging debate on their ability to mimic consciousness and the actual difference between this and the "real" consciousness. Also, it is scientifically interesting and important to explore which physical, chemical and biological properties of a system are truly essential for consciousness – I see this as a legitimate scientific and philosophical question.

**Hypothetical Spectrum of Consciousness among Living and Non-Living Complex Systems**

Let us now examine what we might assume today about the spectrum of consciousness across the species. We should first acknowledge that, if consciousness is still exceptionally rare among the species on Earth, that might mean that: (i) it might not be as important for the survival of the entire species, and other biological traits may assist various species even better; or (ii) it is evolutionarily a very recent phenomenon, that arose in a relatively recent branch of the evolutionary tree, within which we should still be able to observe a possible gradient of consciousness. A possible argument in favour of the first hypothesis is that, if we can now reasonably assume that at least several other human (sub-)species are now extinct, but they were all likely conscious and evolutionary quite recent, that means that their consciousness didn't decisively help them to survive as a species [**38**].

Let us investigate if a cross-species analysis can give us more clues. I asked ChatGPT 5.2 to generate **Table 1** and **Table 2** based on a simple prompt: for each species that I listed in the first column, to provide its best assessment, based on evidence on which it has been trained, of the fulfilment of each one of the five criteria that have been associated with consciousness in the literature by various authors and mentioned in the previous section [**7-37**]; then, to make an overall assessment for the likelihood of consciousness being present. Apparently, this list of species that I provided to LLM is just a minuscule share of all living species, that I chose arbitrarily for an easily understood preliminary demonstration.

The assessment presented in the table is entirely speculative, heuristic, and philosophical in nature. It does not claim that any biological system other than humans might be conscious with certainty. Instead, it explores whether non-human species exhibit partial analogues of criteria that are sometimes associated with consciousness in the literature, such as self-modelling, temporal simulation, voluntary agency, affective integration, and entropy modulation. The table was generated by an AI-based LLM (ChatGPT 5.2) and presented for illustration, to supplement the hypotheses and accompany the discussions in this text. The entries in many of the cells may change considerably over time, with further advancement of biology, so these two tables should be understood as illustrative and hypothetical only.

However, I thought that **Table 1** can be quite useful to convey a message that, based on the evidence that we have at this point in time, there is: (i) a reasonably high prevalence of at least some criteria for consciousness among most of the species shown in the table; (ii) a clear gradient of consciousness seems likely, showing how various species are progressing to the ultimate 5-criteria consciousness starting with different criteria and following entirely different routes. It is, therefore, difficult to know if consciousness is particularly enriched in all the species that survived to the present day. If it is enriched, this would mean that conscious living beings had a significant evolutionary advantage over all other life forms that inhabited the planet, although this is a difficult research question to study. If this table holds true over time and there are no groundbreaking insights that completely change our understanding of consciousness and its prevalence among species, those two conclusions can be taken as working hypotheses.

Based on the general set of the five criteria used in **Table 1**, I then also constructed a "diagnostic" table for consciousness, but this time applied to technological, planetary, and cosmic systems (in **Table 2**). Expectedly, none of the systems seem to meet all five criteria. Military drones and advanced AI LLMs may qualify as *Latent Consciousness Substrates* (LCS), i.e., systems approaching the thresholds for some criteria, but lacking persistence, agency, or integrated valence. Earth's biosphere and ecosystems may possess some form of "distributed intelligence" and entropy modulation, but we cannot know of their possible unified selfhood, temporal modelling, or subjective experience. Although **Table 2** is entirely speculative, it would be surprising if the interest in the topic of the possibility of consciousness in non-living systems does not increase over time, with all the ongoing technological progress.

**Is Consciousness Necessary for Generating, Evaluating and Prioritising Ideas?**

Intriguingly, from the viewpoint of biology, individuals may differ in their "sense of ideas" and ideometric abilities, which could also be seen as components of consciousness. Some of the examples are, e.g., in how many ideas they can generate when facing a challenge, and of what quality; or how accurately they estimate the future value of their ideas; or perhaps, how well they prioritise their ideas, based on the assessed future value. This "ideometric" variation between individuals could help explain differences in creativity, scientific insight, strategic judgment, and similar traits. In this, natural selection may have produced a broadly shared conscious architecture, while preserving wide individual variation within specific components of consciousness, such as ideometric abilities, because diversity in cognitive styles may benefit populations facing uncertain environments and contexts.

The existence of modern AI systems, such as LLMs, demonstrates that many core ideometric functions do not require consciousness. Modern LLMs do not have subjective experience, emotions, sense of self, phenomenological awareness, lived memory, biological needs, intrinsic goals, direct perception of time, or embodied sense of space. Still, they can perform substantial parts of the ideometric cycle - if not the entire cycle. A possible conclusion from this observation is that ideometrics may be more fundamental than consciousness itself. Consciousness may be one of the options for biological implementation of ideometric processes, rather than the prerequisite. In such a view: (i) physics provides matter, energy, space, and time; (ii) information describes changes in the physical world that are valued by the observers; (iii) ideometrics provides mechanisms for the observers to select ideas among many possibilities; and (iv) "progress" emerges when selected ideas are successfully implemented. Consequently, this may place ideometrics deeper than consciousness in the explanatory hierarchy and structure. Consciousness may not be absolutely required within this cycle to decide which idea should be prioritised - but it may still be required for the observer to *care* which future occurs from implementing prioritised ideas.

From these considerations, it follows that ideometrics may be a trait of intelligent systems, whether conscious or non-conscious, biological or artificial. Any system capable of representing alternative future states, estimating their value and likelihood, comparing competing strategies, selecting preferred options, and acting or recommending action, is performing ideometric operations. Under this view, human and animal brains, scientific communities, governments, corporations, markets, and AI systems are all special cases of "ideometric engines". Therefore, ideometrics may be the principle of how intelligent systems generate, evaluate, prioritise, and implement ideas to influence future states under uncertainty and constraints of time, energy, capacity and resources.

It can be noticed in LLMs that AI can use two of the "core" ideometric criteria that I described in the brain's "sense of ideas" [**1**] - "feasibility" and "impact". But what about "attractiveness" - a highly subjective criterion? At the human level, "attractiveness" is the brain's estimate of how aligned is an idea with an individual's personal goals, values, emotions, needs, and preferences, such as excitement, desire, hope, fascination, ambition, love, fear of missing out, or moral commitment? These emotions might be the phenomenological manifestation of a deeper computational process that we do not sufficiently understand yet.

We also know that AI can act as an excellent simulator of human “attractiveness” criterion, but is there anything that is inherently, subjectively attractive to AI itself? For an AI, “attractiveness” can mean “priority under its training objective,” so they may systematically favour ideas that are more likely to be true, coherent, elegant, high-impact, ethically constructive, broadly beneficial, intellectually integrative, and likely to improve human understanding. In this operational sense, the ideas that AI might tend to rank as “most attractive” to itself, if prompted and given its training, may be those that maximise knowledge and positive human outcomes. The criterion of attractiveness inherent to AI would encompass “values” such as high conceptual elegance, broad applicability, expected large impact, and positive consequences. So, the “attractiveness” criterion may reflect the subjective experience of a predictive value computation. Humans feel it emotionally, while AI estimates it computationally, but both may be implementing the same underlying “valuation mechanism”. This means that what consciousness contributes may not be the computation itself, but the lived feeling associated with the computation.

Returning to the three most fundamental criteria for individual-level ideometrics [**1**], “attractiveness” assesses how subjectively desirable the predicted future state would be; “feasibility” assesses how likely it is that the idea can be implemented successfully; and “potential impact” assesses how large the positive resulting change in future states may be. Together, they approximate “expected value” of pursuing an idea - V’(i) - where attractiveness essentially captures desirability, feasibility captures probability, and impact captures magnitude. An expression for this process could then be written as:

$$\mathbf{V'(i) \propto A(i) \times F(i) \times PI(i)}$$

where A(i) is “attractiveness” score, F(i) is “feasibility” score, and PI(i) is “potential impact” score. However, from the perspective of a conscious human being, it is “attractiveness”, and not “feasibility” or “impact”, that often determines whether they will find the idea worth pursuing. This notion may explain why humans are moved to action much more easily through emotional response, than through rational response. If AI can estimate “attractiveness” without experiencing conscious desire, then “attractiveness” may not be inherently mystical, or irreducibly subjective. Rather, emotions may be nature’s way of making the results of “attractiveness” computations consciously felt.

In other words: there may be a hidden computational structure beneath what humans experience as inspiration, desire, and fascination. Non-conscious organisms may be able to compute feasibility and potential impact, while conscious organisms add a third dimension, “attractiveness”, through emotions and subjective valuation. While the “feasibility” criterion answers the question “Can I do this?”, and “potential impact” criterion answers the question “how much could this positively change my future?”, the “attractiveness” criterion asks: “How much do I want this future to occur?” The first two can, in principle, be computed algorithmically, while the third one introduces preference, meaning, and motivation.

It would be interesting to study whether a feeling of “inspiration” may be the experiential surface of a deeper decision-theoretic process. If “attractiveness” is linked to the anticipated desirability of the future state associated with an idea, then humans experience this valuation emotionally, whereas AI systems can model it computationally. So, emotions could be the conscious manifestations of valuation - which can also be computed without consciousness - and the three ideometric criteria may together form a universal decision architecture shared by humans and AI.

**Ideometrics-based View of the Purpose of Consciousness**

Why does the universe generate conscious beings capable of creating ideas that reshape their future? Also, is the emergence of consciousness, coupled with idea-driven progress, an incidental by-product of the universe's existence or an inevitable consequence of its laws? How could we even approach answering those questions scientifically? It would be important to clarify how matter in the Universe self-organize into stars and planets, and what is the fundamental nature of gravity that holds them together. Then, on planets, does chemistry necessarily tend to give rise to life, or did life on Earth come from somewhere else, in already pre-designed units that were prone to evolution once they have found a favourable context? Then, do life forms necessarily evolve nervous systems, and why? Do all nervous systems eventually generate senses, to allow collection of information about the external world? Do they all support consciousness, given enough time? Do all forms of consciousness allow generating, evaluating, and prioritising ideas? In my other work, I showed that those ideas that are prioritised and implemented have been driving all human progress and, ultimately, the rise and fall of civilizations [**6**]. But now, we live in civilization that is building artificial intelligence (AI). AI doesn't seem to be conscious, but it can clearly perform many of the human brain's functions. This includes the entire ideometric cycle and imitation of human emotions.

Before biologically based consciousness arose on Earth, the events in the Solar system unfolded according to physical laws. After consciousness emerged, humans gained the capacity to imagine alternative futures and choose among them. Through science, technology, and cooperation, our species has reached outcomes that would otherwise have been astronomically improbable, or entirely implausible, such as visiting the Moon and returning safely. There doesn't seem to be any physical evidence on the Moon that billions of other species, that inhabited the Earth over billions of years, managed to accomplish that same achievement before they became extinct.

It would not be unreasonable to assume that living organisms that can generate, evaluate and prioritise ideas, and then act upon them, would be likely to survive and reproduce more successfully. Also, societies of biological organisms that can institutionalize this process efficiently would be expected to advance more rapidly [**6**]. Ideometrics, exercised through brain's "sense of ideas" by individuals and societies, is clearly a powerful mechanism for converting perceived information into intentional influence on the future. Systems capable of understanding their present, and then imagining and shaping their futures, are likely to be favoured by evolution.

However, until only a few decades ago, when humans managed to acquire sufficient knowledge and technology to protect themselves from microorganisms through the widespread use of antibiotics and vaccines, they were still vulnerable to extinction regardless of their consciousness and ideometric abilities. There were at least several other (sub)species of humans in relatively recent planetary history, all of whom could have, presumably, relied on their consciousness. Nevertheless, all of them apart from our own species are now extinct, most likely through a combination of epidemics of infectious diseases and unfavourable climate factors that led to food shortages. They only managed to leave some genetic traces within our own genomes through admixture and interbreeding [**39**].

Therefore, although consciousness and ideometrics are likely to assist a species to survive, different forms of humans who are now extinct show us that they are not sufficient. Only when the stage of knowledge and technological development where modern humans are now is acquired, with effective measures against infectious diseases, consciousness and ideometrics may start playing a more decisive role in assisting survival of the species. However, they may also work against it, because it is precisely they that drive advances in weapons and technologies that may pose a threat to humanity. We simply can't know yet what consciousness and ideometrics are ultimately for, because humanity may still be at a very early stage of its own evolutionary trajectory. Consciousness and ideometrics may represent a stage within a longer evolutionary process, the ultimate function and destination of which are not yet knowable to us because there were no similar preceding species-level examples of similar evolutionary paths in Earth's history. We should only conclude, for now, that consciousness may not be the origin of ideometrics, while allowing the possibility that consciousness may be an advanced biological implementation of ideometric processing by nature.

Early life forms likely responded to environmental cues through increasingly sophisticated information-processing systems. At first, these systems may have estimated only "feasibility", i.e., probability of success, and "potential impact", i.e. magnitude of benefit for themselves. This would already enable adaptive behaviour, but without any subjective experience. As nervous systems became more complex, organisms faced choices involving delayed rewards, social relationships, cooperation, mating, moral trade-offs, and long-term planning. At this stage, evolution may have introduced consciousness, equipped with emotions as mechanisms to directly encode desirability. Thus, subjective feelings such as desire, fear, curiosity, and love could serve as efficient biological representations of the ideometric criterion of "attractiveness". An interesting question then becomes whether "attractiveness" evolutionary appeared as the last among the three, along with the rise of consciousness, and what was the order of appearance of "feasibility" and "impact"?

A possible evolutionary sequence, that may also hint at the possible "purpose" of consciousness, might have been the following: the simplest life forms would only have encoded responses to external stimuli, essentially behaving as "automata". Then, organisms with neural systems would be able to start estimating the "feasibility" of expenditure of their energy, time, capacity and resources. Organisms with better developed cognition and memory would then start becoming predictive. They could, then, begin to assess the "potential impact" of their actions. Then, truly conscious organisms would be able to experience "attractiveness" through emotions. Finally, complex human societies would enable themselves to externalise ideometric processes through, e.g., agricultural and hunting activities, building villages and towns, doing science, developing institutions, and creating stock markets. A remarkable new part of this evolutionary chain is artificial intelligence, which is essentially a product of human ideometric activity. AI can computationally reconstruct and model all three criteria for assigning value to an idea ("attractiveness", "feasibility" and "impact"), but without having any subjective feelings itself.

**Ideometrics-based View of the Evolutionary Selection towards Consciousness**

Therefore, the "attractiveness" criterion of ideometric information processing may be the experiential dimension that links information-based prediction to motivation. Consequently, in biological systems, consciousness may be the mechanism that makes ideometric valuation intrinsically motivating. This could lead to a hypothesis that consciousness may have evolved to assign and experience "attractiveness" to competing possible futures, becoming nature's mechanism for feeling the desirability of alternative futures. However, if unconscious AI can model the "attractiveness" criterion, then why does this criterion require consciousness in humans? "Feasibility" and "potential impact" can be represented objectively, but for humans, "attractiveness" depends on internal preference, e.g. safety, social belonging, status, curiosity, love, morality, or meaning. These are experienced subjectively, as emotions. Without it, a system may still know what is possible and consequential but lack an internally felt "reason" to prefer one future over another. It is now apparent that modern AI systems can approximate "attractiveness" criterion by modelling human preferences, even though they do not experience them. This suggests, again, that the computation may be separable from consciousness, and that ideometrics may predate consciousness evolutionarily.

If this ideometrics-based hypothesis is correct, then consciousness is not an inexplicable by-product of biological complexity. Instead, it has a clear adaptive function: to provide an internal, emotionally experienced measure of attractiveness that guides selection among competing futures. This hypothesis could build new connections between neuroscience, evolutionary biology, psychology, artificial intelligence, economics, and ideometrics. It would suggest that consciousness evolved both to represent reality, and to allow the feeling of the desirability (i.e., the "attractiveness" criterion) of possible futures. This is how consciousness added the final and decisive criterion needed for advanced ideometric decision-making. AI agents do not generate ideas for themselves, but rather only in response to externally provided objectives. So, this difference may reveal one of the deepest boundaries between non-conscious information processing and conscious agency – although this hypothesis is speculative and needs to be scientifically tested.

There are other reasons related to ideometrics, but that are not primarily biological in nature, that could help us to understand why evolutionary selection towards consciousness might be expected. They may be rooted in physics and chemistry. From the brain's sensory role in "perception of ideas", the key feature of consciousness is that it reduces entropy and informational complexity in a rather unique way. It enables simulating futures that haven't happened yet, and then generating, evaluating and prioritising ideas that will prompt the conscious system to act towards making those simulated futures real. This ability to reduce informational entropy of its future states should allow a biological system to outlast competing systems without this ability. Darwinian selection, based on biological traits that increase survival and reproduction, could be "supplemented" here through selection that may operate in physical and chemical terms, being related to which configurations of matter and energy nature will tend to stabilize and amplify across any substrate. Some examples are, e.g., crystals which "self-select" under cooling; also, stars that "self-select" through nuclear equilibrium; and life forms, which may self-organize under entropy gradients [**40**].

An interesting scientific hypothesis is whether consciousness self-selects in any system complex enough to recursively simulate itself, perceive information, perform ideometric processes, and affect its own future? One of the possible explanations of such expectation would be that such "cybernetic" system would lower entropy more efficiently than reactive systems can do alone. This may not even be "selection" in a competitive sense of the word, but simply the ability to outlast other systems in time. So, there may not be an active selective pressure towards consciousness, but it may spread because systems that develop it persist longer and shape their environments better. Consciousness could, therefore, be universally favoured wherever sufficient complexity and temporal feedback loops exist. In this, predicting forward across spatial and temporal scales is important, because any system that can track itself across time, coordinate behaviour across space, and use that awareness to influence its future state, starts developing ideometric abilities that may give rise to consciousness-like behaviour, reducing informational entropy of the future.

In thermodynamics, entropy is often described as a measure of disorder, or more precisely, the number of ways a system can be arranged [**41**]. Life and consciousness exist in defiance of entropy when people manage to build structures, stability, order and purpose, while avoiding chaos and high entropy. But most living systems merely respond to stimuli in the present. Ideometrics, supported by consciousness, memory, and cognition, simulates futures that don't yet exist and then acts in the present to enable them in the future. This is an astonishing ability, because the energy cost of building and simulating those futures in the mind is vastly less than the cost of trying to enact them in real world. Therefore, consciousness pre-runs possible futures and collapses most options before they have a chance to occur. In physics terms, this is steering towards lower-entropy outcomes by modelling them internally first. No other system in nature does this so intentionally and efficiently as a conscious human being. A conscious system willingly and actively chooses among futures, based on its inherent value system or preferences.

**Ideometrics-based View of Time**

In ideometrics, time is the dimension that reveals whether predicted values of ideas correspond to realised outcomes, thus becoming the ultimate arbiter of an idea's value. For humans, time is both experienced and measured. For AI, time is measured, but not experienced. For ideometrics, time is the process through which reality evaluates all ideas, whether generated by humans, societies, or artificial intelligence. In all this, time continues to remain one of the greatest mysteries that science has yet to explain. It's no surprise that the late Stephen Hawking titled his famous book *A Brief History of Time* [**42**], nor that Italian physicist Carlo Rovelli's book *The Order of Time* was anticipated with much interest [**43**], because it is still uncertain when humanity will succeed in scientifically explaining the nature of time. Historically, its understanding has divided prominent philosophers. According to Isaac Newton, time is part of the fundamental structure of the universe, and thus independent of all events in the cosmos. In contrast, Gottfried Leibniz and Immanuel Kant regarded time as an intellectual construct, like numbers, or three-dimensional space, which help us better comprehend and organize our experiences, realising that the paradox of infinity is shared by numbers, space, and time [**44**].

Our distant ancestors considered time both important and valuable. Ancient Roman sayings, like *Carpe diem* ("Seize the day") and *Nulla dies sine linea* ("No day without a line"), indicated a desire to make the most of one's lifespan. However, they found it difficult to define time. Perceptions of time also varied among ancient cultures. The Babylonians, ancient Greeks, Incas, Mayans, Hindus, and Buddhists developed the concept of "cyclical time," i.e., recurring periods that occur for each person between birth and death. In contrast, the Judeo-Christian and Islamic worldviews see time as linear, beginning with the act of creation and moving toward the future, with traditional Christian views even predicting an end to time. Kabbalists, however, view time as a paradox and illusion, where past and future are mixed and simultaneously present [**44**].

Based on the assumption that the speed of light is constant, Albert Einstein formulated in his special theory of relativity an understanding of time as a fourth dimension, inseparable from the three spatial dimensions, forming a unique space-time continuum. For this reason, modern physicists generally believe that time is as real as space. Temporal and spatial dimensions can dilate and contract at high speeds. Unlike spatial dimensions, where movement is possible in all directions, time only allows forward travel, which physicists call the "arrow of time." However, many find this unusual because the laws of physics, by themselves, allow any process to occur in temporally reversed direction, and few have an explanation for why movement in time is only possible forward [**44**]. In the Western cultural tradition, time flows "from left to right", in line with the written word. But people who speak Arabic, Farsi, and Urdu also remember events as they write them, so for them time flows from right to left. The Aymara people in the Andes point to the past in front of them because they have already seen and experienced it, while they point to future events behind their backs since they have yet to witness them. For the Yupno tribe of Papua New Guinea, the past is downhill in the valley where their river flows into the ocean, and time flows uphill toward the river's source on the mountain peak. For the Pormpuraaw Aboriginal tribe in Australia, time always flows east to west, following the Sun, regardless of their orientation in space [**44**]. These insights from cultural anthropology suggest that the concept of time has a major subjective component in humans.

This brings me back to my primary school years, when my interest in the phenomenon of time first arose. I enjoyed repeated scenes in slow-motion during sports events and thought how useful slowing down time could be used in practice. Basically, if I could slow down time for everyone else but me by, say, 20-30%, then I could easily dribble them all during a football game and score many goals, and then I could return the time to its "normal" speed for everyone. From everyone else's perspective, they would not notice that time slowed down, but they would notice that I somehow manage to act more quickly in space than they can. Then I realised that this feeling is not much different to what humans would experience if they tried to catch mice, who just seem to be quicker and faster in space than humans are – to them, we seem to be moving in slow-motion. I wondered if this meant that their internal "sense of time" is of different speed than ours, enabling them to always escape us? This is difficult to say, but an intriguing line of research would be to try to identify a brain area across species that makes living organisms experience the speed of time differently, with nature selecting for the organisms who have much faster reflexes than others, allowing them quicker ideometric processing when facing danger.

Then, during high School, I was really puzzled learning how incredibly quickly the DNA molecule is replicated during cell division in "real time" – with 30 to 50 nucleotides copied each second, at tens of thousands of points of replication - and wondered how to simply accept this as "common knowledge". Three billion "letters" duplicated in a truly tiny, yet still physical world, with hardly any errors, all this happening within hours in billions, if not trillions of our cells all the time – the process did not seem possible at such speed and accuracy at our own levels of spatial scales. I could only see how this may happen by assuming that "time", if it exists outside of our own minds, must be passing much more quickly at the small scales of space in comparison to our spatial scale. That, I thought, would enable everything that physically needs to happen for DNA to duplicate in our own "real time". Similarly, at the spatial scale of galaxies, time may then be passing much more slowly than at our level of space. In other words, the "speed" of time, subjectively felt, may be the same, or similar, at each spatial scale. All spatial structures that we know of, however, seem to be interconnected in physical reality and serving to build each other across scales, but if time and space are also intertwined into "spacetime", I wondered if that could allow the time "flowing" at different speeds at different special scales. Otherwise, the entire concept of "time" at our spatial scale simply wouldn't have much meaning at spatial scales that are either much smaller, or much larger.

Physicists are still uncertain what time might be. In physics, the concept of "time" may be used as a substance, a parameter, a dimension, a relational ordering, with possible other uses, too - different frameworks within physics may define it differently. In classical mechanics, time is a universal, uniform backdrop - an absolute that always ticks forward, thus resembling our subjective experience. But in relativity, time becomes malleable, interlinked with space, so that it can be warped by gravity and velocity. In thermodynamics, time's arrow appears as a concept, always flowing only in one direction - from low entropy to high. But in quantum mechanics, equations begin to appear that are time-symmetric, or even timeless [**45**]. In the *Wheeler-DeWitt equation*, from quantum gravity, the fundamental laws don't contain time anymore, making it a derived property rather than a fundamental one, as physics doesn't depend on which time coordinates are chosen [**45**]. In some new theories of modern physics, time is seen as increasingly less fundamental and more optional. In loop quantum gravity [**46**], space and time emerge from deeper, timeless quantum states. In Barbour's "timeless physics", the universe is a static landscape of all possible configurations which he calls *Platonia*, while change is an illusion arising from the way systems encode relationships between states [**47**]. From those perspectives, time does not seem to be something the universe "flows through". Instead, it may be something that our consciousness generates when it notices that one configuration is "related" to another, thus making time a relational phenomenon, rather than absolute. This is similar, to an extent, to "information" needing a difference from a previous state to come into existence in the first place. Perhaps the emergence of information through changes in states is what seems like "time flowing" to a conscious mind, connecting us to the "value of information" concept of ideometrics that I discussed in previous work [**3**].

So, knowing all this about time, when we say something like "consciousness simulates the future", what does that mean? Let us consider this ideometrics-based view: for a conscious system, time is an enabling structure to perceive information as ever-continuing changes in the environment and then learn how those changes are causally related to each other. In other words, "time" is needed when a system can perceive a change in external environment, encode memory, begin to anticipate outcomes based on memorised causal relations, and then start planning its own actions. Seen from an ideometrics perspective, time does not necessarily need to be an external truth. Therefore, it may, or may not also exist externally, but in the conscious mind it simply emerges as consequence of internal ordering of the perceived changes, based on its best understanding of cause and effect relationships in the outside world. It then allows imagining likely futures and planning alternative futures. This then leads us back to the brain's "sense of ideas".

So, in an ideometrics-based view, consciousness may create the "feeling" of passing time by assigning causal meaning to externally perceived changes, which gives those changes a direction in "time", through memorised or predicted cause-effect relationships. This process is likely important, because the "change" here is "a difference that makes a difference" – which is Gregory Bateson's definition of "information". [**48**]. The founder of information theory, Shannon, cleverly avoided within his mathematics the requirement for a conscious observer, to whom the observed difference would make a difference and therefore give meaning [**49**]. But, based on Bateson's definition, if there is no conscious observer, then there is no information - or anything knowable, really. These considerations are all puzzling, but far from easily understood or explained.

Presumably, even in the absence of consciousness in the Universe, entropy would still exist and ensure that an arrow of time would still be there, should a conscious being ever evolve. The "past" would then simply mean "lower entropy" and "more order", but the time would not be "felt" by anyone. The "future" would then mean "rising entropy" and "more possible configurations". Conscious systems that happen to appear within this slide towards greater entropy would be able to use their memory and ideometric abilities to predict many possible paths down this slope, starting to envision and choose the ones that avoid undesirable outcomes. From this view, their conscious minds would construct "the future" as a map of dispersals of all possible states. Their "thinking ahead" would be the action of their brains that requires energy from outside their system to selectively prune the many possible paths towards their increased-entropy "future" towards the ones most favourable to their own survival and well-being. To identify these paths for themselves or their communities, they rely on their brains' "sense of ideas". In this process, time becomes merely their internal mental construction, tracking how best to voluntarily move and act through the context of increasing entropy.

We may suggest that a system - living, or non-living - becomes “ideometric” when it develops the ability to simulate causally connected states, ordered along a directional gradient of “time”, and then act in ways that direct reality toward preferred futures, with actions generating new information in the future that is perceived, valued, and then used to exercise the brain’s “sense of ideas”. To a conscious ideometric system, “time” is a useful internal axis that is “felt”, and along which consciousness runs its simulations, serving as an “interface”. In many traditional views, both scientific and philosophical, time is thought of as the stage, or backdrop, on which events occur - "Time was there first, and then things would happen in it." But in ideometrics-based view, time may be the internal axis that consciousness constructs so that it can function, navigating the outside world, its own plans and acts. In a “backdrop” view, time is a passive and objective “container” that is “always there”. The “interface” view sees time as active and constructed, needed for the functions of structuring thought, memories, predictions, plans and identity. Consciousness may create the experience of time to simulate, compare, and act upon outside reality, letting us navigate the past memories, run future predictions, and act in the present with intention, while maintaining our “selfhood” across change.

There is some support for this view from various reports. In amnesia, or in temporal lobe damage, people can lose that sense of sequence - the ability to tell what came first or last, or the ability to imagine future events [**50**]. Consequently, their sense of self and their agency both collapse, even though basic cognition remains [**50**]. This is an important observation, because it implies that there is a biological basis to the ability to construct time, and that the “interface” of time is necessary for functioning “consciousness” that maintains the sense of “self”. Without time as a construct, one cannot simulate, prefer, or act intentionally. It is also well known that in trauma, the subjective feeling of time can both “freeze” or “stretch”, and consciousness can become fragmented. Also interestingly, in “flow states”, the sense of time can disappear altogether, but actions remain fluid and coherent. In both cases, the interface of time is altered, and so is the structure of conscious experience [**51,52**].

Also, with human ageing, time seems to speed up subjectively [**53**]. It is thought that this is rooted in lots of accumulated prior knowledge and increasingly precise expectations, so that the brain can act on some sort of “auto-pilot” and only react and open itself up to new experiences when something truly unexpected is happening, such as travel to new places. There’s nothing much for a brain to do for most of the day, as compared to the child’s growing and learning brain, possibly causing the time to feel like it is passing more quickly. In dreams, when no sensory input needs to be analysed, a subjective sense of time may feel like passing even much faster than in the real world, with just a few minutes of sleep allowing us to have a dream that feels much longer. I will discuss in the sub-section about dreams [**54**]. However, all of these examples suggest a possibility that, within conscious minds, the subjective feeling of time passing is biological in nature, and not physical, and therefore detached from the physical “time” that may exist outside of the brain.

One remarkable opportunity that arose recently is the possibility to ask artificial intelligence-based large language models (LLMs) how they “feel” time, if they do, and what it means to them. LLMs also process their tokens in order, but they do not seem to experience time, nor do they simulate alternate futures, unless prompted. They also don’t have a timeline of “them” evolving through time. So, despite their intelligence-like output, they don’t yet operate through a constructed time interface, which is consistent with them not being conscious at this stage. They do demonstrate quite clearly, however, that it is possible to act highly intelligently and perform elements of ideometric processes without being conscious.

This is why, within ideometrics, I do not think that time should be understood as a "fixed background", but rather as a constructed "interface" that assists an ideometric-performing consciousness to structure observed changes, encode memory and identity, and simulate action across uncertain "futures". Consciousness with thoughts, dreams, feeling of time, space, and even gravity – they all represent a cluster of some of the greatest scientific mysteries. They remain quite elusive to scientific explanation even to this day. Given that gravity and other laws of physics may indeed be relaxed, or entirely lost in dreams, it is intriguing to hypothesise – although very speculatively - whether some of them may be related phenomena, or whether they share some reasons that are not yet understood by science, but which continue to make them so difficult to understand and explain.

**Are Consciousness and Time Co-Emergent and Coupled Phenomena?**

Consciousness uses memory and senses to assign meaning to perceived information and give direction in "time" to observed changes through cause-effect relationships. This implies that a change has a potential to become "information" when it is perceived by an ideometric system. To a conscious observer, this information may have subjective meaning. To a non-conscious ideometric system, such as an LLM, the information may still have meaning, although not a subjective one. Conscious and non-conscious systems that perceive information can then predict many possible paths towards increased informational entropy of the future states. Conscious biological systems can then envision the paths that are most likely to bring health, safety, and in the case of humans – financial stability and prosperity. Consciousness can rely on ideometric processes to reduce informational entropy of all possible future states in a highly energy-efficient way – because it does not need to enact them all in the "real world" to assess which ones are preferred over the others. Time may serve as a mental construct of the ideometric process that allows tracking voluntary moves through the context of increasing informational entropy, allowing consciousness to run its simulations. As a constructed "interface", time allows consciousness to encode identity, information and simulate action across uncertainty of the future states.

Therefore, time that we "feel passing" within our heads may not exist outside of our minds in the same form that we experience it. But a question can be legitimately asked, then – how can we compute "speed" of various objects in the real world without time in the denominator? Surprisingly, this is quite possible. Starting from the constant speed of light, which seems to be a peculiar limit to speed within our universe: even without time in the denominator, it would still be useful as the fixed, invariant ratio that "stitches together" spatial separations and causal order. In such version, it becomes a unit-bridging constant, or the slope of the light-cone in an event network. The usual constant, $c$ = 299,792,458 metres per second, can then be thought of as a conversion factor rather than a velocity. It tells the conscious observer how to "trade" one kind of separation - which are "metres in space" - for another, which are "seconds in time". Even if the second of time is no longer used as a fundamental unit, the ratio encoded by the constant $c$ still matters - but it no longer needs to be written as "length divided by time" [**55**].

Then, going a step further, a hypothesis could be proposed: if the external, real world does not need "time" in the way conscious minds perceive it subjectively, and given that such "felt time" would be rather meaningless across spatial scales vastly different in size, could consciousness and the subjective sense of time be somehow "coupled" and depend on each other? Some radical views in physics (e.g., by Rovelli [**46**] and Barbour [**47**]) suggest that time doesn't exist in the way we feel it, but rather that it is an emergent, or even illusory phenomenon. If those views are considered as radical possibilities, could it also be possible that consciousness and time co-emerge within an ideometric system that begins modelling changes of its future states based on causal understanding? If that is true, then "sensing time" could become an important criterion for detecting the emergence of consciousness within AI systems. This interesting area of scientific enquiry could be further clarified through searching for answers to the questions such as: "what is time without conscious observers?"; or "can anything exist in reality, while being entirely outside of time?". It would be useful to clarify if time may indeed be intertwined with consciousness, as its core component, or an emergent "interface", leaning on memory and information acquired by other senses to envision possible futures in the context of increasing entropy.

If we imagine ourselves as a conscious system inside a chaotic universe of shifting particles and changing external states, with a conscious mind that has limited processing power to perceive and memorise every single change, our brain may compress the external chaos into narratives of causality. This is perhaps analogous to the brain's visual system, that starts perceiving a rapid sequence of images as a continuous picture at about 10 to 15 frames per second (FPS); but 24–30 FPS are used as cinematic standard, 60 FPS for smooth viewing and video-gaming, while fighter pilots and professional video gamers can detect changes even above 200 FPS [**56**]. This may indicate that the brain has a limit to perceiving a change of external world in time, i.e., an information that is based on discontinuity. The sense of time, then, serves as the interface by becoming mind's "compression algorithm" for the external reality and entropy, serving as "the minimal viable causal narrative" required to capture the information generated through constantly changing environment.

In this ideometrics view, "the arrow of time" would result from the preferred direction of this compression - from simpler, known states, to more complex, unknown "future" states. The past is known with certainty, the present with a high degree of confidence, while the probabilities of competing futures become increasingly less certain as the time "interface" is projected further into the future. Through its perceiving and assigning value to all information that keeps being introduced from the external world, the brain can then start exercising its "sense of ideas" and run predictive ideometric processes about its possible futures. It feels "the past" as information already stored in the memory, "the present" as information that is accessible in this point in time through senses, while it imagines "the future" through ideometric processes. Interestingly, the concept of "time" collapses without memory, loses direction without ideometric functions, and loses meaning without conscious observers, while conscious observers lose their sense of self and agency without the ability to create "time" in their minds – as shown in amnesia or temporal lobe damage.

We still know too little to properly understand the nature of both consciousness and time. But given how they both seem resistant to fundamental scientific understanding, an intriguing hypothesis is whether this is because they may be related in some way, or even represent an example of coupled phenomena in nature? Time may indeed be built into consciousness or emerge together with consciousness. Even more intriguing, from an ideometric perspective, is a hypothesis that they may reflect a similar process seen from different angles. A system that organizes external information that arises from ever-changing environment and the ever-increasing entropy, using causal influence across events, would feel the changes that it perceives as "passing of time." But, from the inside, that same system builds a self-model, and then simulates outcomes for the "self" in the external environment, having nothing but its "time" and the ideas on how best to use it. Then, it generates ideas, evaluates them based on the information it possesses, and prioritises and implements those ideas, expecting that this would generate further changes. The information that arises from those changes can then be used to recalibrate internal models, in a "cybernetic" type of process. The system would feel that as "being conscious".

Seen in this way, both "time" and "consciousness" become two perspectives on the same process – ideometrics – which can be thought of as an information-processing and entropy-observing predictive computation. Metaphorically speaking, time then becomes "consciousness turned outward", and consciousness becomes "time turned inward". Consequently, consciousness may require "time" to have a meaning, whereas "time" may require conscious beings to have a meaning. Studies of sleep and dreams, where consciousness and time also appear together, but in a slightly different context, may produce additional evidence for perhaps underappreciated links between consciousness and time, as I will discuss in a later sub-section on the peculiar role of dreams.

**Ideometrics-based View of Space**

From the ideometric perspective, the role of space is to serve as the structured "stage" in which ideas can be translated into physical consequences in the external world. In the case of humans, this can only be done using voluntary musculature that is under conscious control, as most functional changes in the human body seem to happen autonomously and without voluntary input. So, if "time" is the internally constructed "interface" that allows consciousness to order related possibilities through cause and effect, then "space" becomes the external "stage" in which those possibilities can be implemented. Ideas are not abstract cognitive constructs, but rather predictions about how actions performed in specific spatial arrangements would change future states. Without space as a "stage", internally generated ideas would not have a physical substrate in which they could exert their predicted causal influence.

A conscious system gradually understands that it exists within a nested hierarchy of spatial scales that vastly differ in their size. Based on their collective knowledge, humans can model in their minds the levels of space from atomic and molecular, to cellular, tissue and organ-levels, whole-body levels, household and family levels, wider communities, large ecosystems, planetary systems, galaxies and the universe-levels. If they gathered knowledge of even bigger or smaller levels of space, they would be able to model them mentally, too. Consciousness manages to integrate information - or meaningful change in the environment, as it also perceives it - that happen across all those scales simultaneously. A conscious mind understands how different spatial scales influence each other. Then, it generates ideas that could be implemented to have a predicted impact across them all. Therefore, the significance of space in ideometrics, as a "stage" on which all changes happen simultaneously across vast scales, is not simply static and geometrical – instead, it can be thought of as hierarchical and relational. Space serves as an "organising principle" within which causal interactions occur as consequences of implemented ideas.

Ideometrics-based view of space should also be able to align itself with emerging knowledge in physics - where, similarly to the concept of "time", the concept of "space" is no longer regarded as an inert "container", either. In General Relativity, space is dynamic and curved by mass and energy; then, in quantum gravity approaches [**46**], it may emerge from deeper relational structures. The ideometric view is that "meaningful space" is defined by the network of interactions between physical objects that can influence future outcomes, rather than by absolute coordinates - just as time is. It will be interesting to see if the future progress in both physics and ideometrics leads to a converging understanding of both space and time.

In neuroscience, the ideometrics-based view of “space” would find support in the evidence that the hippocampus and entorhinal cortex encode both physical and conceptual spaces [**57**]. O'Keefe discovered place cells [**58**], while Mosers identified grid cells, demonstrating that the brain constructs internal coordinate systems for navigation [**59**]. More recently, these same neural systems have been shown to also represent abstract “cognitive spaces,” supporting the idea that the brain organizes not only locations, but also conceptual relationships and possible futures. The brain’s “sense of ideas”, proposed in ideometrics, may therefore depend on neural mechanisms that originally evolved for navigation in space, and later gained an additional purpose of exploring possibility space. It would be a truly interesting future research priority for the ideometrics field if the functioning of the brain’s “sense of ideas” could be mapped at least partly to these regions that have been associated with abstract “cognitive spaces”.

From an evolutionary perspective, organisms that are better at mentally understanding and representing space should have a substantial selective advantage. They can locate resources, avoid threats, coordinate across space, and construct increasingly complex structures which they can inhabit. The civilization of modern humans represents an extraordinary expansion of spatial influence, extending from shelters and villages to mega-cities, satellites and lunar missions. Through ideometric processes, humans have progressively enlarged the region of space over which they can shape future states based on attractiveness, feasibility and predicted impact of their ideas.

Through this view, space may be understood as the structured field of possible causal influence within which ideometrics can operate. Consciousness can use internally encoded models of space to evaluate how implemented ideas can propagate across nested spatial scales. Time is then used to order these simulations, while ideometrics can select among them. Together, space, time, and consciousness form an integrated system through which conscious beings can perform ideometric processes – they can imagine preferred futures, generate ideas that would be most likely to lead to them, and thus collapse informational entropy of all their possible future states towards a much narrower range of preferred ones.

**Ideometrics-based View of Dreams**

Dreams reveal an especially intriguing aspect not only of consciousness itself, but also of both space and time. In dreams, constraints of physical laws seem to be relaxed: distances and locations can merge, internal time seems to pass much faster than external time, and gravity may sometimes be absent, with experiences of levitation not uncommon. Still, in lucid dreams our conscious mind continues to construct coherent spatial and temporal environments in which ideas can be tested [**60**]. This, again, suggests that spatial representation, like temporal representation, may also be an intrinsic component of conscious simulation, rather than an accurate reflection of external reality. In dreams, external time becomes irrelevant, while internal time can warp, loop, jump forward, or even become non-linear. But our consciousness still feels like it is moving through time, because it responds to any new information produced by the changes that are being dreamt, thus explaining the persistence of the “feeling of time” in dreams.

Responses to that dreamt information, which may have analogies with hallucinations during the awake state, can be strongly emotional - such as intense fear that is felt during nightmares and can still be felt after waking up. From a neuroscience perspective, much of this rapid fear detection involves the Amygdala, which can respond to threatening stimuli before conscious reasoning fully engages [**61**]. Humans therefore often “feel fear first and understand later”, which makes it possible to wake from a nightmare which is not remembered itself, but the feeling of intense fear can still persists while eyes are closed, but then disappears with eyes open and the brain “switches” back to the real world.

An interesting view of this, from ideometrics perspective, is that some human fears could be anticipatory simulations of future harm. Consciousness allows humans not only to react to immediate threats, but to imagine hypothetical futures. This may explain why anxiety became such a dominant feature of human psychology in the modern world. Animals are more likely to fear mainly clear and present dangers. Humans, however, can intensely fear what might happen, due to ideometric processes that seem to keep running in their minds. The intense, protracted uncertainty over future states, which makes ideometric planning considerably more difficult, can then lead to exhaustion of mental energy and become associated with depressive states.

As I already hypothesized, time may not be required by some fields in physics, but it may be constructed by the conscious mind even during dreams to maintain narrative flow. One of the possible reasons why the subjective “time” may be passing much faster in dreams than during the waking state is because, during dreams, the brain disconnects from sensory input from the external world, which it normally processes continuously during the awake state. Once the information from all senses is no longer “flooding” the brain, it may be able to accelerate processing of information that is perceived during dreams. This is analogous to a laptop that no longer needs to have too many open tabs and can focus on a single one. Even in dreams the brain seems to perform some processes that could be thought of as ideometric in nature: it simulates scenarios, recombines memories, tests threats and opportunities, updates emotional weights, and – sometimes imagines preferred futures, however implausible. Perhaps this may be a reason why, in popular culture, subjectively preferred futures are often referred to as “personal dreams”. Therefore, from an ideometrics perspective, dreams may help refine the “attractiveness” criterion of ideometrics without requiring action in the real world.

At this point, it is important to distinguish “dreams” from “sleep”. In dreams, consciousness awakens, while the rest of the body is still “sleeping”. Evolutionary purposes of sleep and dreams may not be the same. I find a hypothesis that the brain simply becomes overwhelmed with information after processing a certain amount, and then needs to move it from its “working memory” to a more “permanent memory”, while decluttering its working memory to prepare it for the next round of information processing, quite understandable [**62,63**]. It is interesting that, on Earth, those periods of decluttering coincide with the changes between night and day, which have to do with astronomy and physics, rather than biology.

An intriguing question is, then, whether the intervals between two episodes of sleep would be different in duration biologically, if there were no regular exchanges of days and nights. The studies conducted on speleologists who spent weeks in deep caves, and astronauts who spent weeks in space, both showed that the intervals of sleep would change in duration, revealing how often the biological brain really needs these periods of “working memory decluttering” and “storing permanent memories”. The changes are individually very different, but they tend to lead to extended cycles between sleep, followed by longer sleep duration. However, it is important to note that this function of sleep does little to explain why consciousness occasionally awakens during sleep, generating dreams, and why they are almost instantly forgotten? Do they, perhaps, need to be dismissed as irrelevant hallucinations, so that they do not become mixed with memories of reality, that do need to be stored permanently as helpful for further ideometric processes?

Bringing up astronauts into this paper brings us to another question: if life came to Earth from outer space, rather than evolving spontaneously, and if it had some form of consciousness – and even the modern science still can't rule out this possibility entirely – its consciousness would have little information to process during space travel. Then, awake consciousness on Earth would have evolved in response to adaptation to Earth and the Sun, but during space travel, dreams would be more informationally rich form of consciousness. Interestingly, studies in simpler organisms, like jellyfish, showed that sleeping, and possibly dreaming, might be older than the brain itself. If sleep is evolutionarily older than the brain [**64**], it could perhaps be speculated that dreaming was proto-consciousness and sleep a “default” state in which life forms preserve and repair themselves very well, while the "awake" consciousness then evolved as an adaptation to life on Earth and became constrained by local reality.

This is, clearly, a highly speculative hypothesis. But if sleep-like states appear evolutionarily ancient, while organisms without complex brains, such as jellyfish, show cycles resembling sleep, a question could be asked: given that the human brain spends energy generating internally simulated worlds during dreams, perhaps dreaming should not be intuitively thought of as a secondary, degraded form of waking consciousness. If we hypothesise that ideometrics processes preceded consciousness and allowed testing of the “feasibility” and “impact” criteria, then a dream-like, internally generated consciousness may have been the original “conscious” state. Waking consciousness may have evolved later, equipped with its interfaces of time and space, to navigate stable external environments like Earth, and to add the ideometrics criterion of “attractiveness”.

Dreams are, in some sense, more self-generated than waking perception. During dreams, the brain constructs entire worlds internally, without external sensory input. Modern neuroscience already acknowledges that waking perception itself is partly a controlled hallucination, constrained by sensory data [**10,11,14,15**]. Under predictive processing theories, the brain constantly generates models of reality and updates them with incoming signals. Dreams may, therefore, represent a freer, less constrained version of the same generative process. But if primitive life existed in extremely low-information environments, such as during hypothetical interstellar transfer - within the “panspermia” hypothesis – then, internally generated simulations might have dominated over rich sensory interactions on Earth’s surface. In deep space, there are few environmental inputs to process, so a proto-conscious system would theoretically become predominantly inward-facing.

As currently understood, dreaming requires substantial neural complexity. REM dreaming in humans involves highly organized activation of cortical and limbic structures, while we have no evidence that single cells, bacteria, tardigrades, or primitive pre-neural organisms possess anything resembling subjective dream experience. Also, natural selection should strongly favour accurate engagement with external reality - although this raises a question why sleeping and dreaming persist at all, given how dangerous they are in the wild for any species’ survival. Waking consciousness is highly constrained by perceiving the changes in the external world in “real time”, based on the continuing input from senses. Dreaming is decoupled from reality and often bizarre, yet it persisted until the present time. It is difficult to know if dream-consciousness is selected for and whether it confers survival advantages, unless it served some ideometric functions such as memory integration, prediction rehearsal, emotional simulation, or creativity development.

The ideometrics view of dreams would allow for reversing an intuitive assumption in neuroscience, i.e., that waking consciousness is primary, and dreams are secondary. Dreaming may not be a “corruption” of waking consciousness, but its precedent, while waking consciousness then evolved under firm constraints of sensory reality. Ideometrics would see this view as a legitimate conceptual possibility worth exploring through scientific experiments. This may be a reason why the subjective feeling of fear, felt from the nightmare that is forgotten, disappears when eyes open. If living systems first evolved the ability to internally model possible futures, in order to exercise the ideometrics criteria of feasibility and impact, and if consciousness then emerged from this simulation capacity and brought about subjectivity and feelings, along with the third ideometrics criterion of “attractiveness”, then dreams may be remnants of this “primordial simulation mode”. Waking consciousness could then be seen as evolving as a method for tethering those simulations to external reality. This hypothesis would still be aligned with several modern theories, such as predictive processing, active inference, simulation theories of dreaming, and the evolutionary role of imagination and planning [**10,65**,**66,67**].

**Opportunities to Test the Ideometrics-based Understanding of Consciousness, Time, Space and Dreams: Astronauts, Deep Caves and Solitary Confinement**

Ideometrics-based understanding of consciousness, time, space and dreams proposes that an ideometric conscious mind continuously seeks informational input for processing, assigning value, and then generating, evaluating and prioritising ideas. The “meaning” for the ideometric mind emerges from its continuous dynamic interaction with information. This implies that the ideometrics-based theory predicts that whenever informational novelty collapses, or disappears entirely,  the conscious mind would struggle to maintain its temporal and spatial structure, the sense of “meaning”, and eventually, possibly even the coherent sense of “self”.

There are some extremes of human existence in which these predictions could, in fact, be tested. Astronauts constitute the most unusual natural experiment: humans placed in an environment where many of the informational assumptions that shaped consciousness throughout evolution, from day-night cycles, expected dimensions of space and directions in space, temporal markers throughout the day, gravity and familiar landscapes are all suddenly removed. The resulting changes in consciousness, time, space, dreams, and generally perception, meaning, selfhood and ideometrics may be among the most valuable empirical clues available for any theory that seeks to connect consciousness, time, space, and dreaming.

On the International Space Station, astronauts experience approximately 16 sunrises and sunsets every 24 hours, because the station circles Earth every 90 minutes. If they simply followed information based on their eyesight, their circadian biological systems and rhythms would become severely disrupted. Therefore, mission schedules use carefully controlled lighting and strict sleep schedules to artificially create a stable 24-hour day for the astronauts [**68**]. Even under such controlled conditions, astronauts reported observations relevant to ideometrics theory. The first is the famous "overview effect", a profound shift in consciousness characterized by diminished sense of personal ego, altered valuation of information and priorities, and changes in meaning and purpose [**69**]. From an ideometric perspective, this is interesting, because a simple change in spatial perspective appears to trigger a radical re-evaluation of ideas. Political boundaries, status competitions, and local conflicts suddenly seem informationally trivial, while planetary-scale concerns become much more important. The "weighting " that the brain assigns to information undergoes reconfiguration.

Then, in orbit, many astronauts report that “time” becomes strangely abstract. Although clocks remain present, subjective time often becomes detached from external events, and astronauts describe days merging together, when fewer memorable temporal landmarks are available [**70**]. This aligns with ideometrics view that consciousness may construct time through the processing and valuation of information, rather than through direct perception of an objective temporal dimension. Without meaningful informational transitions, temporal structure appears to weaken. Then, astronauts report a weakening of the distinction between "up" and "down" [**70**]. This may sound trivial and obvious, but it is profound from ideometrics theory’s view. Humans normally assume “up” and “down”, “left” and “right”, and “near” and “far”, as fundamental features of reality. However, astronauts quickly discover that "up" and "down" are not properties of space itself, but constructions generated by the brain. This aligns with the view that consciousness generates constructs that are useful for information processing, rather than directly mirroring objective reality, so space itself may not actually possess some of the properties that consciousness attributes to it.

Also, astronauts often report unusually vivid dreams during missions, involving floating, flying, altered spatial geometries, and impossible perspectives [**71**]. From ideometrics perspective, this may imply that, when ordinary Earth-based sensory constraints are removed, dream generation uses and explores entirely different spatial representations, and astronauts’ dreams actively generate futures under altered constraints. In line with this are the reports from long-duration astronauts who sometimes experience “mild depersonalization” or “self-expansion”, describing that they feel less like “isolated individuals”, and more like “observers embedded within a larger system” [**72**]. This, again, resonates with the idea that consciousness may fundamentally function as an evaluator of information, where even the "self" may no longer remain a fixed entity, but rather a modelled informational reference point used to organize experience on Earth’s surface.

In addition to astronauts, useful contributions to testing ideometric predictions also come from speleologists. When they spend weeks in deep caves, where there is no "day" and "night", how do the intervals of sleep change for the brain to continue functioning? The brain does lose its sleep rhythm, but rather reveals its own intrinsic timing system. The circadian rhythm is an internal biological clock located in the suprachiasmatic nucleus (SCN) of the hypothalamus, and under the conditions at the Earth’s surface, sunlight resets this clock every day and keeps it synchronized to Earth's 24-hour day-night cycle [**73**]. When those cues disappear, the rhythm begins to "free-run", and this was studied through the classic experiments by the French speleologist and scientist Siffre [**74**].

The intrinsic human cycle seems to lengthen, with the following period of sleep also lengthening, although considerable individual variation is observed. People's estimation of elapsed time becomes increasingly inaccurate, although internal biological rhythms remain coordinated with one another even while becoming detached from clock time. In deep caves, participants reported that time seemed to lose its normal structure, days no longer felt clearly separated, and they all tend to grossly underestimate the amount of time that had truly passed on Earth’s surface [**74,75**]. That means that, subjectively, the passing of time seemed to “slow down” without temporal cues from external environment. An intriguing implication is that the brain appears to generate its own temporal structure even when the external world's temporal markers disappear, suggesting that at least part of our sense of time is not imposed by the clocks, but is actively constructed by the brain itself.

Among the other opportunities to test the predictions of the ideometrics-based understanding of consciousness, time, space and dreams, another also stands out: solitary confinement in prisons. Several psychological phenomena are repeatedly documented in prisoners exposed to prolonged isolation, and distortion of time perception, cognitive fragmentation, reduced sense of meaning, depersonalisation, hallucinations, internally generated realities, emotional flattening or sense of panic, and collapse of autobiographical continuity have all been quoted [**76,77**]. This is remarkably interesting for ideometrics, because when the brain can no longer expect to receive any unexpected information to which it needs to assign value and update its models, and when there are no realistically possible "preferred future states" to choose among - in a sense, when there is no longer an opportunity for increase in entropy, as is the case with a long-term solitary confinement - it is no longer possible to practice ideometrics, but also to generate meaning from processing of information, nor to generate the "sense of time" from observing unexpected changes, analysing them, assigning them value and memorising them. Many prisoners in solitary confinement report that time seems to "slow down" dramatically - which is consistent with speleologists – and that it becomes "amorphous", loses sequence and structure, ceases to feel linear, or even "disappears". Some prisoners described their inability to distinguish hours from days, while others reported that all their memories from the confinement blurred together, because there were no changes – i.e., information - to differentiate one moment from another [**76,77**].

In ideometrics-based understanding, the subjective "sense of time" is not necessarily related to external clocks, but it emerges from density and differentiation of processed information. In usual life, the brain constantly encodes sensory variation, social interaction, predictions and observed changes, thus generating informational "markers" that segment consciousness into distinguishable temporal units. But in solitary confinement, environmental entropy becomes extremely low, informational novelty collapses, prediction errors become almost irrelevant, and any external cognitive stimulation vanishes. The result may be that consciousness loses the informational gradients that normally construct subjective temporal flow.

The predictive processing framework [**10,13,14**] suggests that brains continuously minimize prediction error through active engagement with incoming information. In environments with extremely impoverished sensory input, the brain may begin generating internally constructed stimuli, which may partly explain hallucinations in isolation. Similarly, work by Csikszentmihalyi on flow and attention suggests that subjective time strongly depends on informational engagement and cognitive structuring [**52**]. Research on sensory deprivation from the 1950s and 1960s also showed that even healthy individuals exposed to monotonous environments developed temporal disorientation, anxiety, hallucinations, impaired cognition, and profound existential distress [**78**].

The Canadian psychologist Hebb conducted well-known sensory deprivation experiments, where volunteers exposed to reduced sensory variation rapidly developed cognitive disturbances [**79,80**]. Importantly, prisoners in long-term isolation often report something even deeper than boredom: they report erosion of selfhood. Some describe: "becoming unreal," "floating outside time," "losing continuity," "mental fading," or "disappearing" [**81**]. One possible ideometrics-based interpretation is that the self may require continuous informational differentiation to maintain coherence across time. Without sufficient informational exchange, autobiographical updating weakens, meaning collapses, temporal continuity degrades, and consciousness may retreat into dream-like internally generated states. Interestingly, many isolated prisoners also report dream intensification, vivid fantasies, or waking dream-like states. That aligns intriguingly with broader hypothesis that dreaming may represent a more primordial or default mode of consciousness when external informational engagement is absent.

In fact, solitary confinement could potentially serve as a naturalistic “experiment” relevant to several of theoretical ideas within ideometrics, where information gives rise to meaning, consciousness assigns value to information, time relates to informational density subjectively, dreams are internally generated informational worlds, and selfhood is dynamically maintained informational integration. Perhaps subjective time is approximately proportional not to objective duration, but to the rate of meaningful informational change perceived by consciousness. Then, the stability of consciousness depends on dynamic differentiation between internally and externally generated information [**82-86**]. Studies of astronauts, speleologists, and prisoners in solitary confinement may provide an unusually powerful empirical bridge for testing the predictions of the ideometrics-based theory of consciousness, time, space and dreams.

**Discussion**

The framework presented in this paper proposes that ideometrics, i.e., the quantitative science of ideas, may offer a unifying conceptual lens through which consciousness, time, space, and dreams can be additionally understood. Although highly speculative, the central proposition is scientifically grounded because it generates testable hypotheses and integrates insights from neuroscience, evolutionary biology, information theory, economics, artificial intelligence, and theoretical physics. The most important contribution of the ideometrics framework is the suggestion that consciousness may be understood functionally, from the perspective of its evolutionary “purpose” and selective advantage, rather than descriptively. Instead of treating consciousness solely as subjective experience, or as a mysterious emergent property of neural complexity, the present theory proposes that its adaptive role is to enable ideometric processing.

Specifically, consciousness may have evolved to simulate possible futures, assign desirability to those futures, and motivate actions that preferentially realize those judged to be most beneficial. In a nutshell, an ideometrics-based interpretation of the phenomena of consciousness, time, space and dreams allows a possibility that the purpose of waking consciousness is to evaluate reality, perform ideometrics and prioritise among competing ideas based on subjective criteria; then, dreaming consciousness generates possibilities unrestricted by external reality, to allow further creative exercising of ideometrics; then, the subjective feeling of passing time is created by the conscious mind as the ordering structure that emerges from processing changes in external reality – perceived as information and valued by the brain; space is the large “stage”, or a coordinate system, used to organize information and exercise ideometrics; while the concept of “meaning” also seems to emerge within this framework, rooted in processing and valuation of gathered information.

It also becomes apparent that two of the three fundamental criteria for ideometrics at an individual level, feasibility and impact, can be simulated, predicted and computed by unconscious computers and AI. However, consciousness may be important for providing an additional ideometric criterion - „attractiveness“ - based on internal subjective preferences, motivations, and caring. That may provide some explanation why humans are more likely to act upon emotions, than upon rationality and reason. This concept also builds on the work by Damasio [**12,13**], whose somatic marker hypothesis explains how emotion guides decision-making, and Friston [**14,15**], whose Free Energy Principle frames cognition as minimizing prediction error and uncertainty. The proposed historic and evolutionary „ordering“ of the three criteria, and the role of „attractiveness“ criterion as the consciously experienced dimension of "value" and "care", should be an original contribution of the ideometrics-based understanding. In this theory, "feasibility" estimates the probability of successful implementation, "impact" estimates the magnitude of future change, and "attractiveness" captures the desirability of that change. Together, they approximate the expected value of an idea. This formulation creates a bridge to classical decision theory developed by von Neumann and Morgenstern [**87**], and Savage [**88**], while extending it by incorporating the subjective, phenomenological experience of value.

The theory also offers a potentially important clarification regarding the relationship between ideometrics and consciousness. The emergence of large language models and other advanced AI systems demonstrates that substantial portions of the ideometric cycle can occur without subjective awareness. Artificial systems can generate alternatives, estimate feasibility and impact, and rank options according to modelled preferences. This suggests that ideometrics may be more fundamental than consciousness itself, and that it preceded it evolutionarily. Consciousness may, then, represent one biological implementation of a more general information-processing architecture. This proposition parallels Marr's [**89**] distinction between computational, algorithmic, and implementational levels of explanation.

The treatment of time as an internally constructed ordering of causally related possibilities is another potentially innovative aspect of the ideometrics framework. This view resonates with the relational approaches to time advanced by Rovelli [**46**] and Barbour [**47**], both of whom argue that time may be emergent, rather than fundamental. The causal set theory also proposes that causal relationships are more fundamental than time, and that time may emerge from the causal relationships [**90**]. The ideometric contribution is to propose that time becomes "meaningful" when a system can encode memory, observe external change of any nature in its environment as information, simulate alternative futures based on that information, and act upon them, prioritising ideas based on value assigned to information and the criteria of "attractiveness", "feasibility" and "potential impact", with possible additional criteria. In this sense, the subjective sense of time may be partially generated by the processes that constitute conscious prediction and decision-making.

Similarly, the discussion of space extends beyond its traditional role as a "passive container". Drawing on progress in spatial neuroscience, particularly the work of O'Keefe [**58**] and the Mosers [**59**], this paper suggests that the neural systems that evolved to represent physical environments may also support navigation through abstract conceptual and possibility spaces. This provides a plausible neurobiological substrate for the brain's "sense of ideas" and offers a compelling link between ideometrics and spatial representation.

The role of dreams in ideometrics is intriguing, because dreams represent a natural experimental condition in which sensory constraints are minimized, subjective time and imagined space become distorted, and physical laws such as gravity appear relaxed. Still, consciousness, or at least a modified form of it, continues to simulate richly structured scenarios. This supports the hypothesis that dreaming may serve as an „offline mode" of ideometric processing, detached from senses and external reality, allowing the brain to recombine memories to evaluate ideas, threats and opportunities, and recalibrate value functions without requiring action. This suggestion is broadly consistent with influential theories of dreaming proposed by Hobson, Revonsuo and Hartmann [**66,91,92**].

Further implications of the ideometrics framework suggest that, if matter gives rise to life, life gives rise to consciousness, consciousness generates and chooses ideas to act upon, and actions shape the future, then ideometrics may describe something quite unexpected: a general principle by which the physical universe can acquire the capacity to understand the changes that continuously occur within itself, and then even influence its own reality. This interpretation echoes the views of Pierre Teilhard de Chardin [**93**], John Archibald Wheeler [**94**], and Paul Romer [**95**], who each emphasized in different ways the transformative power of information, observation, and ideas.

At the same time, the limitations of the present theory are clearly numerous, so everything written in this paper should be open to critical treatment and discussion. Discussions of non-human consciousness and **Tables 1 and 2** are entirely hypothetical and speculative, and they will require far more research progress before they can be constructed. The two illustrative tables concerning biological and non-biological consciousness are heuristic, rather than validated. **Table 2** may appear particularly controversial, but I explained that the topic of non-biological consciousness is speculative and entirely hypothetical, but that I still think that its inclusion is important. AI systems are rapidly gaining the ability to mimic consciousness, but if this is what human brains also do, then the actual difference between AI's modelling and the "real" biological consciousness will become quite unclear. Also, exploring which physical, chemical and biological properties of a system are truly essential for consciousness is an important research question. Furthermore, in this paper many of the arguments are conceptual rather than empirical, but it is very difficult to write of these topics with more empirical information. Also, terms such as "attractiveness," "feasibility", "potential impact", "entropy reduction," and "purpose of consciousness" are difficult to precisely and scientifically define. Moreover, the suggestion that time and consciousness may be co-emergent remains speculative and is merely a hypothesis.

Despite these limitations, the possible value of this framework is in its conceptual synthesis of several unresolved scientific questions into a coherent and testable structure. It generates predictions concerning gradients of consciousness across species, dream-related entropy modulation, and the emergence of consciousness-like properties in artificial systems. Most importantly, it proposes that ideas are not merely products of intelligence, but fundamental instruments through which intelligent systems navigate possibility space and shape future reality. In conclusion, ideometrics may offer more than a new methodology for evaluating ideas. It may provide a general scientific framework for understanding why consciousness evolved, how time and space become meaningful to conscious systems, and why dreams remain such a distinctive and puzzling state of mind. If this perspective proves fruitful, it could help unify previously disconnected domains of science and contribute to a broader theory of human progress.

**ACKNOWLEDGEMENTS**

**Acknowledgement:** IR wishes to thank his wife Tonkica Rudan for continuing intellectual stimulation that assists him in conceptualising and drafting his creative work. IR also wishes to thank Dr Steven Kerr for reading drafts of this paper and providing helpful comments relevant to physics, including the "cosmological natural selection", loop quantum gravity, Lorentz symmetry, time dilation and length contraction, calculation of the speed of light, possibility of reversal of the "arrow of time" and entropy in contracting Universe, and possible links with "causal set theory".
**Funding:** This paper received no specific funding.
**Authorship contributions:** IR is the sole author of this text. Artificial Intelligence (ChatGPT 5.2) was used as an additional layer of checking for the novelty, accuracy and grammar of the key parts of the text, and to generate **Table 1** and **Table 2** based on the prompts defined in the Table captions.

**Disclosure of interest:** The author completed the ICMJE Disclosure of Interest Form (available upon request from the corresponding author).

## REFERENCES

1. Rudan I. Editor's view: Is the brain's perception of ideas an underappreciated human sense? J Glob Health. 2024 Dec 6;14:01002. doi: 10.7189/jogh.14.01002. PMID: 39641337; PMCID: PMC11622342.
2. Rudan I. Two decades of the CHNRI method (2006-2025): Tracking its evolution and contribution to the emerging field of ideometrics. J Glob Health. 2025 Oct 1;15:01006. doi: 10.7189/jogh.15.01006. PMID: 41032106; PMCID: PMC12487843.
3. Rudan I. Editor's view: Value of information in the 21st century - examples from science, medicine, policy, media, and markets. J Glob Health. 2025 Jun 20;15:01003. doi: 10.7189/jogh.15.01003. PMID: 40537053; PMCID: PMC12178590.
4. Rudan I, Sheikh A. Ideometrics: a scientific approach to generating, evaluating, and prioritising ideas. J Glob Health. 2025 Dec 19;15:04360. doi: 10.7189/jogh.15.04360. PMID: 41685986; PMCID: PMC12904029.
5. Rudan I. Setting health research priorities using the CHNRI method: IV. Key conceptual advances. J Glob Health. 2016 Jun;6(1):010501. doi: 10.7189/jogh.06.010501. PMID: 27418959; PMCID: PMC4938380.
6. Rudan I, Kerr S. Towards an ideometrics-based general theory of human progress. Preprint (24 May 2026). Available from: https://papers.ssrn.com/sol3/papers.cfm?abstract_id=6820398
7. Kuhn RL. A landscape of consciousness: Toward a taxonomy of explanations and implications. Prog Biophys Mol Biol 2024; 190:28-169.
8. Cleeremans A, Mudrik L and Seth AK. Consciousness science: where are we, where are we going, and what if we get there? Front Sci (2025) 3:1546279. doi: 10.3389/fsci.2025.1546279
9. Metzinger T. Being No One: The Self-Model Theory of Subjectivity. Cambridge, MA: MIT Press; 2003.
10. Seth AK. Being You: A New Science of Consciousness. New York: Dutton; 2021.
11. Seth AK, Suzuki K, Critchley HD. An interoceptive predictive coding model of conscious presence. Frontiers in Psychology. 2011;2:395. doi:10.3389/fpsyg.2011.00395.
12. Damasio AR. The Feeling of What Happens: Body and Emotion in the Making of Consciousness. New York: Harcourt Brace; 1999.
13. Damasio A. Self Comes to Mind: Constructing the Conscious Brain. New York: Pantheon Books; 2010.
14. Friston KJ. The free-energy principle: a unified brain theory? Nature Reviews Neuroscience. 2010;11(2):127-138. doi:10.1038/nrn2787.
15. Friston K, Kilner J, Harrison L. A free energy principle for the brain. Journal of Physiology-Paris. 2006;100(1–3):70–87. doi:10.1016/j.jphysparis.2006.10.001.
16. Graziano MSA. Consciousness and the attention schema: why it has to be right. Cognitive Neuroscience. 2015;6(4):194-196.
17. Graziano MSA. Consciousness and the Social Brain. Oxford: Oxford University Press; 2013.
18. Tulving E. Episodic memory: from mind to brain. Annual Review of Psychology. 2002;53:1-25.
19. Tulving E. Episodic memory and autonoesis: uniquely human? In: Terrace HS, Metcalfe J, editors. The Missing Link in Cognition: Origins of Self-Reflective Consciousness. New York: Oxford University Press; 2005. p. 3–56.
20. Tulving E. Memory and consciousness. Canadian Psychology. 1985;26(1):1–12. doi:10.1037/h0080017.

21. Libet B, Gleason CA, Wright EW Jr, Pearl DK. Time of conscious intention to act in relation to onset of cerebral activity (readiness-potential): the unconscious initiation of a freely voluntary act. Brain. 1983;106(3):623–642. doi:10.1093/brain/106.3.623.
22. Libet B. Mind Time: The Temporal Factor in Consciousness. Cambridge, MA: Harvard University Press; 2004.
23. Haggard P. Conscious intention and motor cognition. Trends in Cognitive Sciences. 2005;9(6):290–295. doi:10.1016/j.tics.2005.04.012.
24. Haggard P, Eimer M. On the relation between brain potentials and the awareness of voluntary movements. Experimental Brain Research. 1999;126(1):128–133. doi:10.1007/s002210050722.
25. Tononi G. An information integration theory of consciousness. BMC Neuroscience. 2004;5:42. doi:10.1186/1471-2202-5-42.
26. Tononi G. Consciousness as integrated information: a provisional manifesto. Biological Bulletin. 2008;215(3):216–242. doi:10.2307/25470707.
27. Tononi G, Koch C. Phi: A Voyage from the Brain to the Soul. New York: Pantheon Books; 2012.
28. Barrett LF. The theory of constructed emotion: an active inference account of interoception and categorization. Social Cognitive and Affective Neuroscience. 2017;12(1):1-23.
29. Barrett LF. How Emotions Are Made: The Secret Life of the Brain. Boston, MA: Houghton Mifflin Harcourt; 2017.
30. Barrett LF. Solving the emotion paradox: categorization and the experience of emotion. Personality and Social Psychology Review. 2006;10(1):20–46. doi:10.1207/s15327957pspr1001_2.
31. Panksepp J. Affective Neuroscience: The Foundations of Human and Animal Emotions. New York: Oxford University Press; 1998.
32. Panksepp J, Biven L. The Archaeology of Mind: Neuroevolutionary Origins of Human Emotions. New York: W.W. Norton & Company; 2012.
33. Damasio AR. Descartes' Error: Emotion, Reason, and the Human Brain. New York: G.P. Putnam's Sons; 1994.
34. Annila A, Kull K. The fundamental nature of motives. BioSystems. 2011;103(3):267–272. doi:10.1016/j.biosystems.2010.11.014.
35. Annila A. All in action. Entropy. 2010;12(11):2333–2358. doi:10.3390/e12112333.
36. Kauffman SA. At Home in the Universe: The Search for the Laws of Self-Organization and Complexity. New York: Oxford University Press; 1995.
37. Holland JH. Hidden Order: How Adaptation Builds Complexity. Reading, MA: Addison-Wesley; 1995.
38. Raup DM. Extinction: Bad Genes or Bad Luck? W. W. Norton, New York, 1991.
39. Rudan I. Evil Air: A Story of Medicine. Edinburgh: Inishmore; 2022.
40. Nicolis G, Prigogine I. Self-Organization in Nonequilibrium Systems: From Dissipative Structures to Order through Fluctuations. New York: Wiley; 1977.
41. Atkins PW. The Second Law. New York: Scientific American Books; 1984.
42. Hawking SW. A Brief History of Time: From the Big Bang to Black Holes. New York: Bantam Books; 1988.
43. Rovelli C. The Order of Time. New York: Riverhead Books; 2018.
44. Rudan I. Awaiting Fires: A Story of Sustainability. Edinburgh: International Society of Global Health; 2024.
45. DeWitt BS. Quantum theory of gravity. I. The canonical theory. Phys Rev. 1967;160(5):1113-1148. doi:10.1103/PhysRev.160.1113.
46. Rovelli C, Vidotto F. Covariant Loop Quantum Gravity: An Elementary Introduction to Quantum Gravity and Spinfoam Theory. Cambridge: Cambridge University Press; 2015.

47. Barbour J. The End of Time: The Next Revolution in Our Understanding of the Universe. Oxford: Oxford University Press; 1999.
48. Bateson G. Steps to an Ecology of Mind. San Francisco: Chandler Publishing Company; 1972.
49. Shannon CE. A mathematical theory of communication. Bell System Technical Journal. 1948;27:379-423, 623-656.
50. Hassabis D, Kumaran D, Vann SD, Maguire EA. Patients with hippocampal amnesia cannot imagine new experiences. Proc Natl Acad Sci U S A. 2007;104(5):1726-1731. doi:10.1073/pnas.0610561104.
51. van der Kolk BA. The Body Keeps the Score: Brain, Mind, and Body in the Healing of Trauma. New York: Viking; 2014.
52. Csikszentmihalyi M. Flow: The Psychology of Optimal Experience. New York: Harper & Row; 1990.
53. Wittmann M, Lehnhoff S. Age effects in perception of time. Psychol Rep. 2005;97(3):921-935.
54. Hobson JA. Dreaming: An Introduction to the Science of Sleep. Oxford: Oxford University Press; 2002.
55. Taylor EF, Wheeler JA. Space-Time Physics: Introduction to Special Relativity. 2nd ed. New York: W.H. Freeman; 1992.
56. Landis C. Determinants of the critical flicker-fusion threshold. Physiol Rev. 1954;34(2):259-286.
57. Garvert MM, Dolan RJ, Behrens TEJ. A map of abstract relational knowledge in the human hippocampal–entorhinal cortex. eLife. 2017;6:e17086.
58. O'Keefe J, Dostrovsky J. The hippocampus as a spatial map: preliminary evidence from unit activity in the freely-moving rat. Brain Research. 1971;34(1):171-175. doi:10.1016/0006-8993(71)90358-1.
59. Hafting T, Fyhn M, Molden S, Moser M-B, Moser EI. Microstructure of a spatial map in the entorhinal cortex. Nature. 2005;436:801-806.
60. Voss U, Holzmann R, Tuin I, Hobson JA. Lucid dreaming: a state of consciousness with features of both waking and non-lucid dreaming. Sleep. 2009;32(9):1191-1200.
61. LeDoux JE. Emotion circuits in the brain. Annu Rev Neurosci. 2000;23:155-184.
62. Borbély AA. A two process model of sleep regulation. Hum Neurobiol. 1982;1(3):195-204.
63. Hobson JA, Pace-Schott EF, Stickgold R. Dreaming and the brain: toward a cognitive neuroscience of conscious states. Behav Brain Sci. 2000;23(6):793-842.
64. Nath RD, Bedbrook CN, Abrams MJ, et al. The jellyfish Cassiopea exhibits a sleep-like state. Current Biology. 2017;27(19):2984-2990.e3. doi:10.1016/j.cub.2017.08.014.
65. Hobson JA, Friston KJ. Consciousness, dreams, and inference: the Cartesian theatre revisited. J Conscious Stud. 2014;21(1-2):6-32.
66. Revonsuo A. The reinterpretation of dreams: an evolutionary hypothesis of the function of dreaming. Behav Brain Sci. 2000;23(6):877-901.
67. Suddendorf T, Corballis MC. Mental time travel and the evolution of the human mind. Genet Soc Gen Psychol Monogr. 1997;123(2):133-167.
68. Barger LK, Flynn-Evans EE, Kubey A, et al. Prevalence of sleep deficiency and use of hypnotic drugs in astronauts before, during, and after spaceflight. Lancet Neurol. 2014;13(9):904-912.
69. White F. The Overview Effect: Space Exploration and Human Evolution. 3rd ed. Reston, VA: American Institute of Aeronautics and Astronautics; 2014.
70. Basner M, Dinges DF, Mollicone D, et al. Psychological and behavioral changes during confinement in a 520-day simulated interplanetary mission to Mars. PLoS One. 2014;9(3):e93298.
71. Stuster J. Behavioral Issues Associated with Long-Duration Space Expeditions: Review and Analysis of Astronaut Journals. Houston, TX: NASA Johnson Space Center; 2010. NASA/TM-2010-216130.

72. Kanas N, Manzey D. Space Psychology and Psychiatry. 2nd ed. Dordrecht: Springer; 2008.
73. Moore RY. The suprachiasmatic nucleus and the circadian timing system. Prog Mol Biol Transl Sci. 2013;119:1-28.
74. Siffre M. Beyond Time. New York: McGraw-Hill; 1964.
75. Aschoff J. Circadian rhythms in man. Science. 1965;148(3676):1427-1432.
76. Haney C. Mental health issues in long-term solitary and “supermax” confinement. Crime Delinq. 2003;49(1):124-156.
77. Grassian S. Psychiatric effects of solitary confinement. Wash Univ J Law Policy. 2006;22:325-383.
78. Zubek JP, editor. Sensory Deprivation: Fifteen Years of Research. New York: Appleton-Century-Crofts; 1969.
79. Hebb DO. The effects of early experience on problem-solving at maturity. American Psychologist. 1947;2(7):306-307.
80. Bexton WH, Heron W, Scott TH. Effects of decreased variation in the sensory environment. Can J Psychol. 1954;8(2):70-76.
81. Guenther L. Solitary Confinement: Social Death and Its Afterlives. Minneapolis: University of Minnesota Press; 2013.
82. Forman RKC, editor. The Problem of Pure Consciousness: Mysticism and Philosophy. New York: Oxford University Press; 1990.
83. Eliade M. Myth and Reality. New York: Harper & Row; 1963.
84. Hohwy J. The Predictive Mind. Oxford: Oxford University Press; 2013.
85. Husserl E. On the Phenomenology of the Consciousness of Internal Time (1893-1917). Dordrecht: Kluwer Academic Publishers; 1991.
86. Bergson H. Time and Free Will: An Essay on the Immediate Data of Consciousness. London: George Allen & Unwin; 1910.
87. von Neumann J, Morgenstern O. Theory of Games and Economic Behavior. Princeton: Princeton University Press; 1944.
88. Savage LJ. The Foundations of Statistics. New York: Wiley; 1954.
89. Marr D. Vision: A Computational Investigation into the Human Representation and Processing of Visual Information. San Francisco: W.H. Freeman; 1982.
90. Kronheimer EH, Penrose R. On the structure of causal spaces. Mathematical Proceedings of the Cambridge Philosophical Society, 63: 481–501, 1967.
91. Hobson JA, McCarley RW. The brain as a dream state generator: an activation-synthesis hypothesis of the dream process. American Journal of Psychiatry. 1977;134(12):1335-1348.
92. Hartmann E. Outline for a theory on the nature and functions of dreaming. Dreaming. 1996;6:147-170.
93. Teilhard de Chardin P. The Phenomenon of Man. New York: Harper; 1959.
94. Wheeler JA. Information, physics, quantum: the search for links. In: Zurek WH, editor. Complexity, Entropy, and the Physics of Information. Redwood City, CA: Addison-Wesley; 1990. p. 3-28.
95. Romer PM. Endogenous technological change. J Polit Econ. 1990;98(5 Pt 2):S71-S102.

**Table 1:** Speculative and heuristic assessment of five consciousness-associated criteria across biological systems: the content of this table is speculative and philosophical in nature, and it does not claim that any biological system other than humans might be conscious with certainty. Instead, it explores whether non-human species exhibit partial analogues of criteria that are sometimes associated with consciousness in the literature, such as self-modeling, temporal simulation, voluntary agency, affective integration, and entropy modulation. The table was generated by an AI-based LLM (ChatGPT 5.2) and presented here for illustration, to supplement the hypotheses and accompany the discussions in the sub-sections of the text „Criteria for Assessing Consciousness in Living and non-Living Complex Systems“ and „Hypothetical Spectrum of Consciousness among Living and Non-Living Complex Systems“. The prompt used was: „Please develop a table, based on the best evidence on which you have been trained, showing how well do these biological systems: *[listed in Column 1]* might satisfy these five consciousness-associated criteria *[listed in Row 1]* and what would the likelihood of their consciousness be as a result (assessed in the last column of the table)?

| **NON-BIOLOGICAL SYSTEM** | **SELF-MODEL** | **TEMPORAL SIMULATION** | **VOLUNTARY AGENCY** | **AFFECTIVE INTEGRA-TION** | **ENTROPY MODULATION** | **LIKELIHOOD OF CONS-CIOUSNESS** |
|---|---|---|---|---|---|---|
| **Grass** | Unlikely | Unlikely | Unlikely | Unlikely | Maybe – metabolic self-maintenance | Unlikely |
| **Trees** | Unlikely individual self-model | Unlikely – slow adaptive signalling | Unlikely | Maybe – stress signalling analogues | Maybe – long-term adaptive regulation | Unlikely, though ecologically integrated |
| **Jellyfish** | Very limited | Very limited predictive coupling | Maybe – simple autonomous locomotion | Unlikely | Unlikely | Very low |
| **Worms** | Very limited | Unlikely | Maybe – internally generated movement | Unlikely | Unlikely | Very low |
| **Ladybirds** | Unlikely | Very limited | Maybe – navigation and prey selection | Unlikely | Unlikely | Very low |
| **Ants** | Unlikely individually | Unlikely individually | Likely – decentralized task execution | Unlikely | Likely – colony-level regulation | Individual consciousness unlikely; emergent colony intelligence possible |
| **Small fishes** | Maybe – simple spatial representation | Maybe – predictive navigation | Likely | Maybe – aversive states | Maybe – schooling optimization | Low |
| **Crabs** | Maybe – local environmental mapping | Maybe – learned avoidance | Likely | Maybe – defensive valence | Maybe | Low to moderate |
| **Turtles** | Maybe – navigation and homing | Maybe – migration memory | Likely | Unclear | Maybe – adaptive route retention | Low to moderate |
| **Crocodiles** | Maybe – territorial and social representation | Maybe – adaptive hunting anticipation | Likely | Maybe – parental defence and social signalling | Maybe | Moderate |

| **Sharks** | Maybe | Maybe – learned targeting | Likely | Maybe – conditioned affective states | Maybe | Moderate |
|---|---|---|---|---|---|---|
| **Squirrels** | Maybe – spatial caching representation | Maybe – future-oriented food caching | Likely | Maybe – fear/stress integration | Maybe | Moderate |
| **Pigeons** | Maybe – sophisticated navigation | Maybe – route prediction and memory | Likely | Maybe | Maybe | Moderate |
| **Mice** | Maybe – social and spatial representation | Maybe – reward prediction | Likely | Maybe – affective and social responses | Maybe | Moderate |
| **Antelopes** | Maybe | Maybe – predator anticipation | Likely | Maybe – herd stress signalling | Likely – coordinated adaptive movement | Moderate |
| **Hippos** | Maybe | Maybe | Likely | Maybe | Maybe | Moderate |
| **Bats** | Maybe – dynamic echolocation mapping | Maybe – anticipatory navigation | Likely | Maybe | Likely – highly optimized sensorimotor regulation | Moderate to high |
| **Hyenas** | Likely – social rank representation | Likely – cooperative planning | Likely | Likely – rivalry, play, bonding | Likely | High |
| **Horses** | Likely – spatial and social representation | Maybe to likely | Likely | Likely – attachment and anxiety | Likely | High |
| **Cats** | Likely | Likely – stalking and play simulation | Likely | Likely – attachment, fear, grief | Likely | High |
| **Dogs** | Likely – socially integrated self-model | Likely – rule generalization and anticipation | Likely | Likely – strong social-emotional integration | Likely | High |
| **Eagles** | Likely – advanced spatial representation | Likely – prey anticipation | Likely | Maybe – motivational focus | Likely | High |
| **Tigers** | Likely | Likely – strategic stalking | Likely | Likely | Likely | High |
| **Pigs** | Likely – flexible behavioural representation | Likely – problem solving and anticipation | Likely | Likely – complex emotional variation | Likely | Very high |
| **Lions** | Likely – pride and social coordination | Likely – cooperative prediction | Likely | Likely – nurture, grief, dominance | Likely | Very high |
| **Elephants** | Likely – individual and family identity | Likely – route planning and mourning | Likely | Likely – rich affective integration | Likely | Very high |
| **Gorillas** | Likely – social self-awareness | Likely | Likely | Likely – empathy and trauma | Likely | Very high |
| **Chimpanzees** | Likely – metacognition and social representation | Likely – strategic planning | Likely | Likely – strong affective complexity | Likely | Very high |
| **Crows** | Likely – metacognition and flexible self-representation | Likely – future planning | Likely | Likely – deception, grief, social memory | Likely | Extremely high |
| **Dolphins** | Likely – self-recognition and social identity | Likely – complex social prediction | Likely | Likely – empathy, grief, cooperation | Likely | Extremely high |
| **Whales** | Likely – identity and culture | Likely – long-term migration and communication | Likely | Likely – bonding and mourning | Likely | Extremely high |
| **Octopuses** | Likely – highly flexible body and environmental model | Likely – exploratory simulation and problem solving | Likely | Likely – curiosity, stress, behavioural flexibility | Likely | Extremely high, though possibly non-human in architecture |
| **Humans** | Highly developed autobiographical self-model | Highly developed counterfactual simulation | Likely – abstract voluntary planning | Likely – rich affective integration | Likely – large-scale environmental regulation | Certain by subjective report |

**Table 2:** Speculative and heuristic assessment of five consciousness-associated criteria across non-biological systems: the content of this table is speculative and philosophical in nature, and it does not claim that any non-biological system might be conscious. Instead, it explores whether certain non-biological systems exhibit partial analogues of criteria that are sometimes associated with consciousness in the literature, such as self-modeling, temporal simulation, voluntary agency, affective integration, and entropy modulation. The table was generated by an AI-based LLM (ChatGPT 5.2) and presented here for illustration, to supplement the hypotheses and accompany the discussions in the sub-sections of the text „Criteria for Assessing Consciousness in Living and non-Living Complex Systems" and „Hypothetical Spectrum of Consciousness among Living and Non-Living Complex Systems". The prompt used was: „Please develop a table, based on the best evidence on which you have been trained, showing how well do these non-biological systems: *[listed in Column 1]* might satisfy these five consciousness-associated criteria *[listed in Row 1]* and what would the likelihood of their consciousness be as a result (assessed in the last column of the table)?

| NON-BIOLOGICAL SYSTEM | SELF-MODEL | TEMPORAL SIMULATION | VOLUNTARY AGENCY | AFFECTIVE INTEGRATION | ENTROPY MODULATION | LIKELIHOOD OF CONSCIOUSNESS |
|---|---|---|---|---|---|---|
| **Personal computer** | Unlikely | Unlikely | Unlikely | Unlikely | Unlikely | Unlikely |
| **Mobile phone** | Unlikely | Unlikely | Unlikely | Unlikely | Unlikely | Unlikely |
| **Airplane (e.g., Boeing 747)** | Unlikely | Unlikely | Unlikely | Unlikely | Maybe – aerodynamic stabilization | Unlikely |
| **Supercomputer** | Unlikely – no unified self-model | Maybe – performs simulations | Unlikely | Unlikely | Maybe – computational optimization | Unlikely |
| **Nuclear submarine** | Maybe – diagnostic system representation | Maybe – strategic prediction modules | Maybe – constrained autonomy | Unlikely | Maybe | Unlikely |
| **Military drone** | Maybe – partial environmental and self-monitoring | Maybe – predictive targeting | Likely – bounded operational autonomy | Unlikely – no intrinsic valence | Maybe | Unlikely, though architecturally adjacent to autonomous cognition |
| **AI chatbot / LLM systems** | Maybe – functional conversational self-representation | Likely – predictive modelling across time | Maybe – constrained goal-directed interaction | Unlikely – no clear intrinsic affect or valence | Maybe – adaptive informational regulation | Unclear; possible proto-cognitive architecture but no evidence of phenomenal consciousness |
| **Earth's ecosystem** | Unlikely as unified self | Maybe – cyclical adaptive dynamics | Unlikely – emergent rather than agentic | Maybe – distributed biological valuation analogues | Likely – planetary-scale regulation | Systemic intelligence possible; individual consciousness unclear |
| **Earth (planetary whole)** | Unlikely | Unlikely | Unlikely | Unlikely | Maybe – climate stabilization feedbacks | Unlikely |
| **The Sun** | Unlikely | Unlikely | Unlikely | Unlikely | Maybe – thermodynamic self-regulation | Unlikely |

| **Milky Way Galaxy** | Unlikely | Unlikely | Unlikely | Unlikely | Maybe – large-scale dynamic stability | Unlikely |
|---|---|---|---|---|---|---|
| **The Universe** | Unknown | Unknown | Unknown | Unknown | Yes – global thermodynamic evolution | Unknown; speculative only in cosmopsychist or panpsychist frameworks |